\documentclass[american,uppersorbian,english]{article}
\usepackage{bm}%
\usepackage[colorlinks=true,linkcolor=blue]{hyperref}%
\usepackage[T1]{fontenc}
\usepackage{algorithm}
\usepackage[font={small,it}]{caption}
\usepackage{authblk}
\usepackage{float}
\usepackage{mathtools}
\usepackage{amsmath}
\usepackage{amsthm}
\usepackage{amssymb}
\usepackage{graphicx}
\usepackage{algpseudocode}
\usepackage{soul} 
\usepackage{subfig} 
\usepackage{esint}
\usepackage{xfrac}
\usepackage[normalem]{ulem} %to strike the words
\usepackage[font={small}]{caption}

\theoremstyle{remark}
\newtheorem{rem}{\protect\remarkname}
\providecommand{\remarkname}{Remark}

\begin{document}
\hyphenation{title}
\author[1]{Mattia Serra\thanks{Email address for correspondence: serram@ethz.ch}}
\author[2]{Pratik Sathe\thanks{pratik\_sathe@iitb.ac.in}}
\author[3]{Francisco Beron-Vera\thanks{fberon@rsmas.miami.edu}}
\author[1]{George Haller\thanks{georgehaller@ethz.ch}}
\affil[1]{\textit{Institute for Mechanical Systems, ETH Z\"{u}rich, Leonhardstrasse 21, \\8092 Zurich, Switzerland}}
\affil[2]{\textit{Department of Physics, Indian Institute of Technology-Bombay, India}}
\affil[3]{\textit{Rosenstiel School of Marine and Atmospheric Science, University of Miami, 4600 Rickenbacker Cswy., Miami, FL 33149, USA}}

\title{Uncovering the Edge of the Polar Vortex}%

\maketitle

\date{}

\begin{abstract}
 The polar vortices play a crucial role in the formation of the ozone hole and can cause severe weather anomalies. Their boundaries, known as the vortex `edges', are typically identified via methods that are either frame-dependent or return non-material structures, and hence are unsuitable for assessing material transport barriers. Using two-dimensional velocity data on isentropic surfaces in the northern hemisphere, we show that elliptic Lagrangian Coherent Structures (LCSs) identify the correct outermost material surface dividing the coherent vortex core from the surrounding incoherent surf zone. Despite the purely kinematic construction of LCSs, we find a remarkable contrast in temperature and ozone concentration across the identified vortex boundary. We also show that potential vorticity-based methods, despite their simplicity, misidentify the correct extent of the vortex edge. Finally, exploiting the shrinkage of the vortex at various isentropic levels, we observe a trend in the magnitude of vertical motion inside the vortex which is consistent with previous results.
	
\end{abstract}

\section{Introduction}

 The formation, deformation and break-down of the polar vortices play an influential role in stratospheric circulation, both in the northern and the southern hemispheres. The beginning of winter is characterized by a rise in circumpolar wind velocities in the stratosphere, resulting in a vortical motion delineated by a transport barrier that isolates polar air from the tropical stratospheric air. This coherent air mass is commonly referred to as the main vortex, while its enclosing barrier is known as the vortex `edge' {\cite{mizuta2001chaotic,nash1996}.\par 

 The spatial extent and strength of the vortex edge determine the severity, location and size of the ozone hole in the stratosphere \cite{WMO1999,McIntyre1995,schoeberl1991,shephard2007}, which is more prominent in the southern hemisphere. The chemical composition of the polar stratospheric air differs significantly from the mid-latitudinal air \cite{levoy1985,profitt1989,loewenstein1989,russell1993}, and the dynamics of the polar vortex profoundly affect such composition \cite{olascoaga2012,huck2005,hood2005}. 
 The chemical isolation and low temperatures inside the polar vortex are necessary for the formation of Polar Stratospheric Clouds (PSCs), the main factor responsible for the rapid stratospheric ozone loss during spring.\par
 
 The material deformation of the vortex edge also exerts a major influence on Earth's surface weather \cite{mitchell2013influence}. Except during vortex break-off events, the cold air in the vortex-interior remains well-isolated from its exterior \cite{schoeberl1989,schoeberl1992,hartmann1989,hartmann1989b,mcintyre1989}. Therefore, the exact spatial extent of the polar vortex edge, its material deformation and the vertical motion of air within the vortex \cite{rosenfield1994} are crucial for understanding the vortex dynamics and its impact on earth's climate, ozone hole variability and the composition of stratospheric air.
 \par

\begin{figure}[h]
	\hfill{}\subfloat[]{\includegraphics[height=0.4\columnwidth]{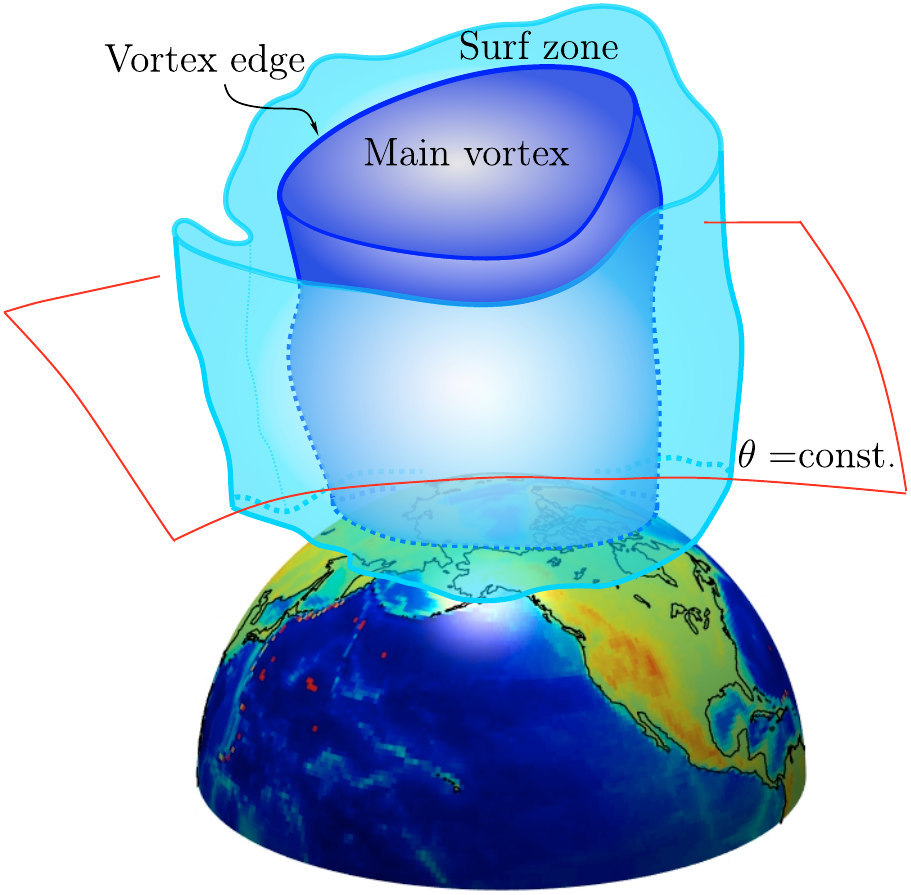}\label{fig:3dIllustr}}\hfill{}\hfill{}\hfill{}\hfill{}
	\subfloat[]{\includegraphics[height=0.35\columnwidth]{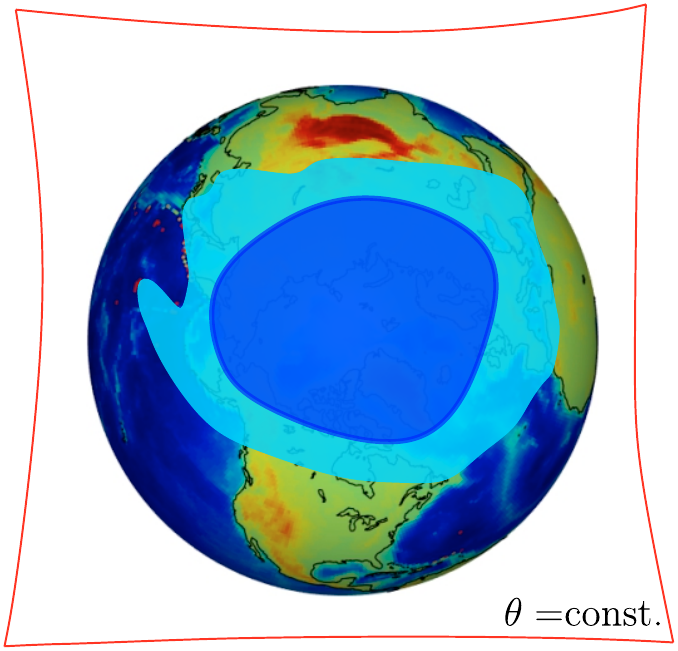}\label{fig:2dIllustr}}\hfill{}
		\caption{(a) Sketch of the polar vortex and its two dynamically distinct regions: the coherent main vortex enclosed by the vortex edge, and the incoherent surrounding surf zone. The main vortex retains cold air while the highly mixing surf zone sheds filaments of warmer air to lower latitudes. (b) Cross section of the polar vortex on an isentropic surface ($\theta=$const.), where $\theta$ denotes the potential temperature.}
	\label{fig:surf_zone}
\end{figure}

\begin{rem}
Broadly used methods developed for locating the polar vortex edge agree that the polar vortex consists of two distinct regions: (a) the main vortex which represents the coherent polar vortex core, and (b) the surf zone which is a wide surrounding incoherent region interacting with lower latitudes (cf. Fig. \ref{fig:surf_zone}). To quote directly from the seminal paper by Juckes and McIntyre \cite{juckes1987} "\textit{...remind us of the likely importance of considering the main vortex as a material entity or isolated airmass, for dynamical as well as chemical purposes."}
\end{rem}

All this suggests the presence of a distinguished  material surface confining the coherent evolution of the main vortex from the incoherent pattern of the surf zone. However, the most popular approaches to locating the vortex edge are Eulerian, returning non-material structures based on non-material coherence principles. We summarize these approaches below.

From a meteorological perspective, heuristic methods for assessing the shape and movement of the polar vortex are based on the area diagnostic \cite{butchart1986,baldwin1988climatology} and later extensions, termed elliptic diagnostics \cite{waugh1997,waugh1999,hannachi2011}, which are mainly based on Potential Vorticity (PV). The vortex edge is then routinely identified by locating steep gradients of PV \cite{mcintyre1984,hood2005,nash1996,steinhorst2005}. Inspired by these heuristics, Nash et. al. \cite{nash1996} define the vortex edge as the closed PV contour with the highest gradient relative to the area it encloses. Other popular methods include effective diffusivity \cite{nakamura1996,haynes2000,hauchecorne2002,Allen2001} and minimally-stretching PV contours \cite{chen1994alone,norton1994,waugh1994CAS}. From a chemical perspective, various other heuristic methods have been developed based on the chemical composition of the vortex-interior \cite{McDonald2013,Krutzmann2008,Sparling2000}.\par

Lagrangian methods have also been used to locate the polar vortex edge.
Bowman \cite{bowman1993a} was the first to use the Finite Time Lyapunov Exponent (FTLE) for studying the Antarctic  polar vortex. \textcolor{black}{Beron-Vera et al. \cite{beron2012} identify zonal jets in the lower stratosphere as transport barriers using locally minimizing curves of FTLE fields}. \textcolor{black}{ Several papers describe the connection between meridional transport barriers and stratospheric zonal jets \cite{beron2012, beron2008zonal, olascoaga2012, rypina2006, beron2010invariant} often based on identifying trenches of FTLE.} De la Camara et al. \cite{camara2010} investigate the austral spring vortex breakup in 2005 based on high FTLE values, while Lekien and Ross \cite{lekien2010} develop FTLE computation technique over non-Euclidean manifolds, and use it to identify the 2002 Antarctic polar vortex breakup. Koh and Legras \cite{koh2002} compute the Finite Size Lyapunov Exponent (FSLE) field over 500 K isentropic data from the European Centre for Medium-Range Weather Forecasts (ECMWF) to identify transport barriers. Joseph and Legras \cite{joseph2002} conclude that high values of FSLE do not correctly identify the vortex boundary, but rather delineate a highly mixing region outside the boundary called the `stochastic layer'. Other Lagrangian analyses are based on optimally coherent sets \cite{santitissadeekorn2010} and trajectory length \cite{madrid2009,camara2012a,smith2014}. \textcolor{black}{These methods, however, either lack objectivity (frame-invariance), or identify non-material structures. Furthermore, none of them give a rigorous parameter-free definition of the vortex edge but rather a visual indication of its approximate location through steep gradients of appropriate scalar fields.}\par

Here we apply the recently developed mathematical theory of geodesic LCS to locate the polar vortex edge. \textcolor{black}{LCSs are distinguished surfaces in a dynamical system that invariably create coherent trajectory patterns over a finite-time interval \cite{review}.} This kinematic (model-independent) theory has already been applied to detect coherent structures in various atmospheric and oceanic flows \cite{shearless,blackholes,jupiter,automated,Serra2016}. Specifically, elliptic LCSs are frame-independent distinguished material surfaces that attain exceptionally low deformation over a finite-time interval \cite{review}. This physical principle seems to be tailored precisely to identify the polar vortex edge (cf. \textit{Remark 1} and Fig. \ref{fig:surf_zone}).\par

We compute elliptic LCSs on various isentropic surfaces in the northern middle stratosphere in late December 2013 and early January 2014, when an exceptional cold weather was recorded in the northeast United States. We show that a geodesic vortex boundary, defined as the outermost elliptic LCS around the polar vortex region, forms an optimal, non-filamenting transport barrier dividing the vortex core from the surf zone, and hence identifies the exact theoretical location of the polar vortex edge. We prove this optimality by materially advecting the geodesic vortex boundary along with a family of small perturbations to this boundary. Remarkably, despite the smallness of the perturbations, we find all of them to undergo significant deformations while the geodesic vortex boundary shows no filamentation.\par

We also find that in late December 2013, the polar vortex edge is initially centered and undeformed, while in the early January 2014, it materially deforms towards the Northeastern US, consistently with the severe weather phenomena recorded over that time period. The deformation of the polar vortex is typically visualized  through the non-material evolution of the PV field. We show, however, that the PV field misidentifies the correct extent of the vortex edge.
Finally, we calculate the change in the area enclosed by the vortex edge with time, on various isentropic levels. This indicates a trend in the magnitude of vertical motion inside the vortex with respect to the altitude, in agreement with the results of \cite{rosenfield1994}.

\section{Geodesic LCS theory}
\subsection{Set-up and notation}
Consider a two-dimensional unsteady velocity field
\begin{equation}
\dot{x}=v(x,t),\quad x\in U \subset \mathbb{R}^2, \quad t\in [t_0,t_1].
\end{equation}
where $U$ denotes a flow domain of interest. The trajectories of fluid elements define a flow map
\begin{equation}
F_{t_0}^t(x_0):=x(t;t_0,x_0),
\end{equation}
which takes every point $x_0$ at time $t_0$ to its position $F_{t_0}^t(x_0)$ at time $t$.
The right Cauchy-Green strain tensor is often used to characterize Lagrangian strain generated by the flow map, defined as \cite{truss}
\begin{equation}
C_{t_0}^t(x_0)=[\nabla F_{t_0}^t(x_0)]^T \nabla F_{t_0}^t (x_0), \label{eq: CG definition}
\end{equation}
where $\nabla F_{t_0}^t $ is the gradient of the flow map and and $T$ denotes matrix transposition. The tensor $C_{t_0}^t$ is symmetric and positive definite, with eigenvalues $0<\lambda_1\leq\lambda_2$ and an orthogonal eigenbasis $\{\xi_1,\xi_2\}$, which satisfy
\begin{equation}
\label{eqn:eqlabel}
C_{t_{0}}^{t}(x_{0})\lambda_{i}(x_{0})=\lambda_{i}(x_{0})\xi_{i}(x_{0}),\qquad\left|\xi_{i}\right|=1,\ \ i=1,2; \quad\xi_{2}= \left(\begin{array}{cc}
0 & -1\\
1 & 0 \end{array}\right)\xi_{1}.
\end{equation}
A common diagnostic for hyperbolic (i.e., attracting and repelling) LCSs in the flow is the FTLE field $\Lambda_{t_0}^t$ \cite{review}, defined as
\begin{equation}
\Lambda_{t_0}^t(x_0)= \frac{1}{t-t_0}\log \sqrt{\lambda_2(x_0)} \label{eq: FTLE}.
\end{equation}
The FTLE measures the maximum separation exponent of initially close particles over the time interval $[t_0,t]$. Its high values provide an intuitive idea of the location of most repelling material lines in the flow. The FTLE, however, does not carry information about the type of deformation causing particle separation, and hence the underlying LCSs need further post-processing to be identified reliably \cite{haller2002lagrangian}.\par
A more precise variational approach classifies LCSs into three different types depending on the distinguished impact they have on nearby deformation patterns.
Specifically, initial positions of hyperbolic LCSs (generalized stable and unstable manifolds),
elliptic LCSs (generalized KAM tori) and parabolic LCSs (generalized jet cores) are computable as solutions of specific variational principles \cite{review,serra2016efficient}. Later positions of these LCSs are then obtained by advecting their initial positions under the
flow map. We now briefly recap the theory of elliptic LCSs used in this paper. \par

\subsection{Elliptic LCSs}
A typical set of fluid particles deforms significantly when advected under the flow map $F_{t_0}^{t}(\cdot)$. One may seek coherent material vortices as atypical sets of fluid trajectories that defy this trend by preserving their overall shape. These shapes should be bounded by closed material lines that rotate and translate, but show no appreciable stretching or folding.
Motivated by this observation, Haller and Beron-Vera \cite{blackholes} seek Lagrangian vortex boundaries as the outermost closed material lines across which the averaged material stretching shows no leading-order variability.\par
Mathematically, consider a closed smooth curve $\gamma \subset U$, parametrized in the form $x(s)$ by its
arclength $s\in [0, \sigma]$. 
The averaged material strain along $\gamma$, computed between the times $t_{0}$ and $t$, is given by \cite{blackholes}
\begin{equation}
Q=\frac{1}{\sigma} \int_0^\sigma \sqrt{\frac{\langle x'(s),C_{t_0}^t (x(s))x'(s)\rangle }{\langle x'(s),x'(s)\rangle}} \mathrm{d}s, 
\end{equation}
where $\langle \cdot,\cdot \rangle$ denotes the Eulerian inner product and $(\cdot)'$ denotes differentiation with respect to $s$. By the smoothness of the velocity field, $\mathcal{O}(\epsilon)$ perturbations to the material line
$\gamma$ will typically lead to $\mathcal{O}(\epsilon)$ variability in the averaged tangential stretching $Q$ \cite{arnold}. Elliptic LCSs, in contrast, are sought as exceptional closed material lines whose $\mathcal{O}(\epsilon)$ perturbations show $\mathcal{O}(\epsilon^2)$ variability in the averaged tangential stretching, i.e., $\delta Q=0$.

Haller and Beron-Vera \cite{blackholes} show that closed material lines $\gamma$ satisfying $\delta Q(\gamma)=0$ coincide with closed null-geodesics of the Lorentzian metric  
\begin{equation}
g_{x,\lambda}:=\langle x', E_\lambda(x) x'\rangle, \ \ \ E_\lambda(x)=\frac{1}{2} [C_{t_0}^t(x_0)-\lambda ^2I],
\end{equation}
for some constant $\lambda > 0$.
Adopting recently developed results for a fully-automated computation of closed null-geodesics \cite{serra2016efficient}, we compute the initial (time-$t_0$) position of elliptic LCSs ($\gamma$) as the $x-$projection of closed orbits of the initial value problem family 
\begin{equation}
\begin{aligned}
x^{\prime} & = e_{\phi}:=[\cos\phi,\sin\phi]^T, \\
\phi^{\prime} & = -\frac{\cos^{2}\phi\langle\nabla_{x}C^{11}(x),e_{\phi}\rangle+\sin 2\phi\langle\nabla_{x}C^{12}(x),e_{\phi}\rangle+\sin^{2}\phi\langle\nabla_{x}C^{22}(x),e_{\phi}\rangle}{\sin 2\phi [C^{22}(x)-C^{11}(x)]+2\cos 2\phi C^{12}(x)},\\
x(0) & =\left\{ x \in U:\ \ C^{11}(x)-\lambda^{2}=0\right\},\ \ \ \phi(0)=0. \label{eq:ellLCSs}
\end{aligned}
\end{equation}
Here $\phi$ denotes the angle enclosed by the $x^\prime$ direction and the horizontal axis, and $C^{ij}(x)$ denotes the entry at row $i$ and column $j$ of the matrix $C_{t_{0}}^{t}(x)$. Elliptic LCSs are closed curves whose arbitrarily small subsets stretch uniformly by the same factor of $\lambda$. The time-$t$ positions of these LCSs can be obtained by advecting their initial position under the flow map $F_{t_0}^t(\gamma)$. Geodesic vortex boundaries can then be identified as the outermost elliptic LCSs computed over a set of lambdas ranging from the weakly contracting ($\lambda<1$) to the weakly stretching ($\lambda >1$) curves. 

\section{Data and numerical methods}

Synoptic scale stratospheric mixing is quasi-layer-wise and stratified, hence, most  of the analyses are done on isentropic surfaces. For purely adiabatic flows, atmospheric motion is restricted to isentropic surfaces, an approximation which holds true over a period of 7-10 days in the stratosphere \cite{morris1995,morris2002}, consistently with the time interval considered in our analysis.

Using isentropic surface data from the ECMWF global reanalysis \cite{erainterim}, we compute elliptic LCSs on the sequence of isentropic levels $850, 700, 600, 530$ and $475$K using a range of $\lambda$-values from 0.8 to 1.2 in 0.1 steps. The wind velocity and potential vorticity on each isentropic surface are available on a $0.75^\circ$ latitude $\times 0.75 ^\circ$ longitude mesh-grid, with a time resolution of 6 hours. We focus on the time period from $28^{th}$ December 2013 to $8^{th}$ January 2014, when exceptionally severe cold weather affected the northeastern coast in the US. 

Our region of interest is the northern polar region. In a standard spherical coordinate system (cf. Fig. \ref{fig:coordinates1}), the zonal (longitudinal) wind velocity (in $deg./day$) tends to infinity at the north pole, making trajectory calculations intractable in its vicinity. To this end, using a change of coordinates, we shift this singularity to a different location, far from the region we intend to analyze (cf. Fig. \ref{fig:coordinates2}).

\begin{figure}[h]
	\hfill{}\subfloat[]{\includegraphics[width=0.3\columnwidth]{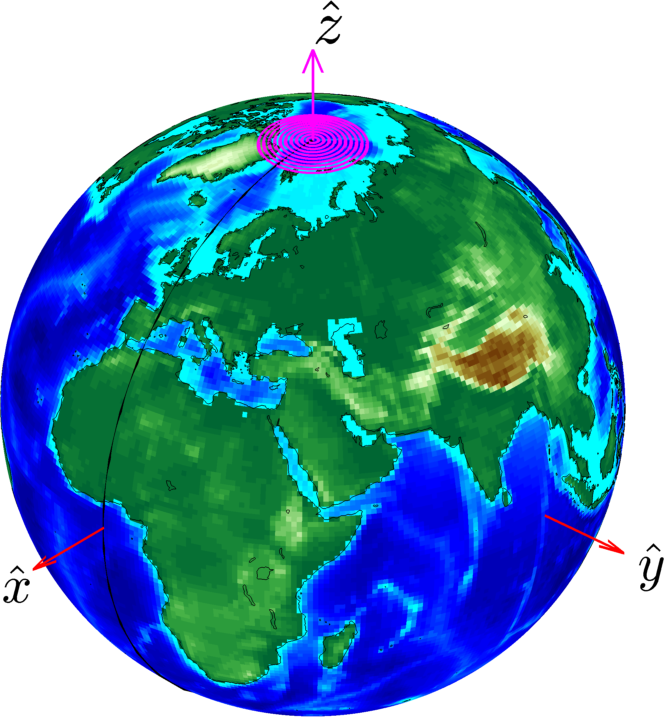}\label{fig:coordinates1}}\hfill{}\hfill{}\hfill{}
	\subfloat[]{\includegraphics[width=0.35\columnwidth]{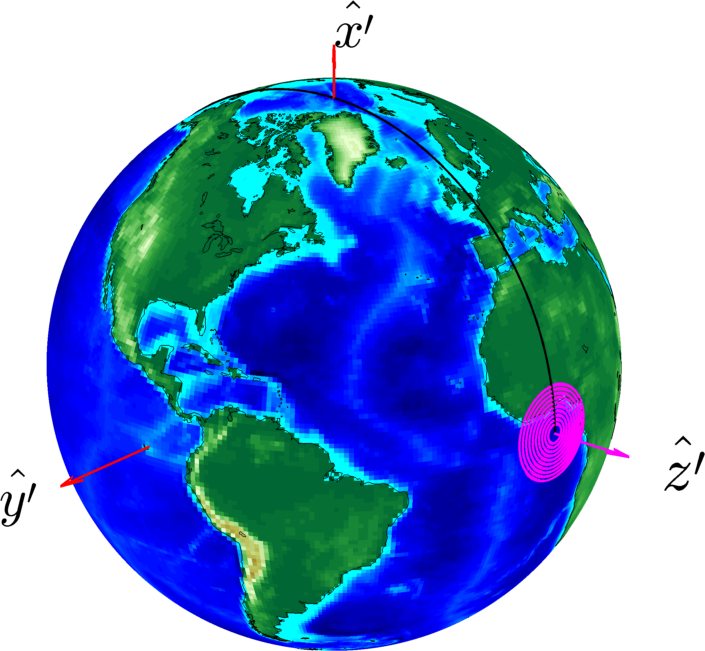}\label{fig:coordinates2}}\hfill{}
	\caption{(a) Original coordinate system. (b) New of coordinate system suitable for Lagrangian analysis close to the North Pole. Magenta circles mark the computationally intractable region for trajectory calculations.}
	\label{fig:change of coordinates}
\end{figure}

We use cubic interpolation of the velocity field for trajectory integration and compute $\nabla F_{t_0}^t(x_0)$ according to \cite{distinguished_surfaces}. Specifically, we compute trajectories of 4 initial conditions (on an auxiliary grid) surrounding every point of the main grid, with the auxiliary grid size equal to $1\%$ of the main grid size. We carry out the trajectory integration and the LCS calculations from \eqref{eq:ellLCSs} using ODE45 in MATLAB, with absolute and relative tolerances set to $10^{-6}$.\par

\section{Results}

\subsection{Elliptic LCSs identify the polar vortex edge}
We perform a backward time analysis over a time interval $[t_0,t_1]$ of 10 days, with $t_0=7^{th}$ January 2014 and $t_1=28^{th}$ December 2013. 
From the positions of the outermost elliptic LCSs (i.e., geodesic vortex boundaries) on the five isentropic surfaces, we construct a 3D visualization of the geodesic polar vortex edge spanning the middle and lower stratosphere (cf. Fig. \ref{fig:funnels}). Such a 3D visualization enables us to make conclusions about the overall material deformation of the main vortex over the time interval we analyze. Specifically, Fig. \ref{fig:funnel_isen_89} shows the vortex edge on 28$^{th}$ Dec. 2013, which has an almost undeformed circular shape, while Fig. \ref{fig:funnel_isen_99} shows the deformed vortex edge on $7^{th}$ Jan. 2014. 

\begin{figure}[h]
	\hfill{}\subfloat[]{\includegraphics[width=0.36\columnwidth]{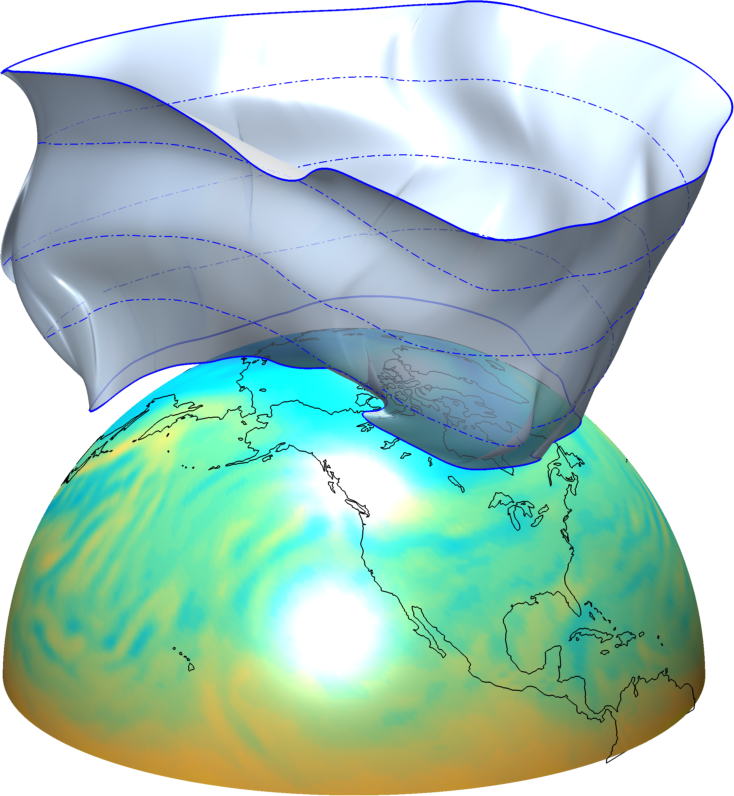}\label{fig:funnel_isen_89}}\hfill{}
	\subfloat[]{\includegraphics[width=0.34\columnwidth]{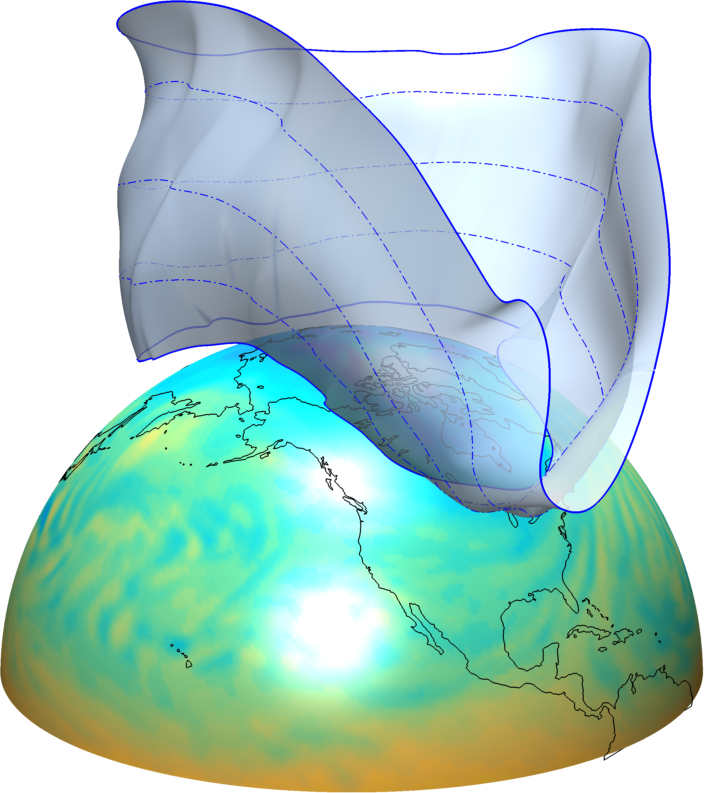}\label{fig:funnel_isen_99}}\hfill{}
	\caption{Reconstructed 3D visualization of geodesic polar vortex edges obtained on all isentropic surfaces under consideration, (a) on $28^{th}$ Dec. 2013 and (b) on $7^{th}$ Jan. 2014. (Figures not to scale along the radial direction). PV on the $475$ K isentropic surface is plotted over the Earth's surface. The full evolution on the vortex boundary over the 10-day time window is available here \href{https://www.dropbox.com/s/9d1ykd9egktry0r/Figure3MultimediaView.mp4?dl=0}{Video 1}.}
	\label{fig:funnels}
\end{figure}

Note that the vortex edge deforms towards the northeastern coast of the US, consistent with the exceptional cold recorded there over the early January 2014. A video showing the evolution of the 3D geodesic vortex edge over the 10-days time window is available here \href{https://www.dropbox.com/s/9d1ykd9egktry0r/Figure3MultimediaView.mp4?dl=0}{Video 1}. 

\subsubsection{Optimal coherent transport barrier} \label{subsubsection:optimality}
Recall that the vortex edge is thought as the outermost closed material line dividing the main vortex from the surf zone (cf. \textit{Remark 1}). Here we show the optimality of the geodesic vortex boundary obtained as the outermost elliptic LCS. The optimal boundary of a coherent vortex can be defined as a closed material surface that encloses the largest possible area around the vortex and undergoes no filamentation over the observational time period. 
To this end, we consider a class of small normal perturbations to the geodesic vortex boundaries. The amount of perturbation ranges between $2^\circ$ and $5^\circ$ (i.e., 5\% to 10\% of the equivalent diameter of the outermost elliptic LCSs). We then advect the geodesic vortex boundary and its perturbations over the time window under study.\par

As an illustration, Fig. \ref{fig:600K_bands} shows the initial positions of the geodesic vortex edge and its perturbations on the $600$K isentropic surface. Figures \ref{fig:475K_optimal}-\ref{fig:850K_optimal}, instead, show the advected images of the geodesic vortex and their perturbations on different isentropic surfaces. The geodesic vortex edge remains coherent under advection in all cases, in sharp contrast to the perturbations which undergo substantial filamentation over the $10$-day time window.

\begin{figure}[h]
	\subfloat[]{\includegraphics[width=0.31\columnwidth]{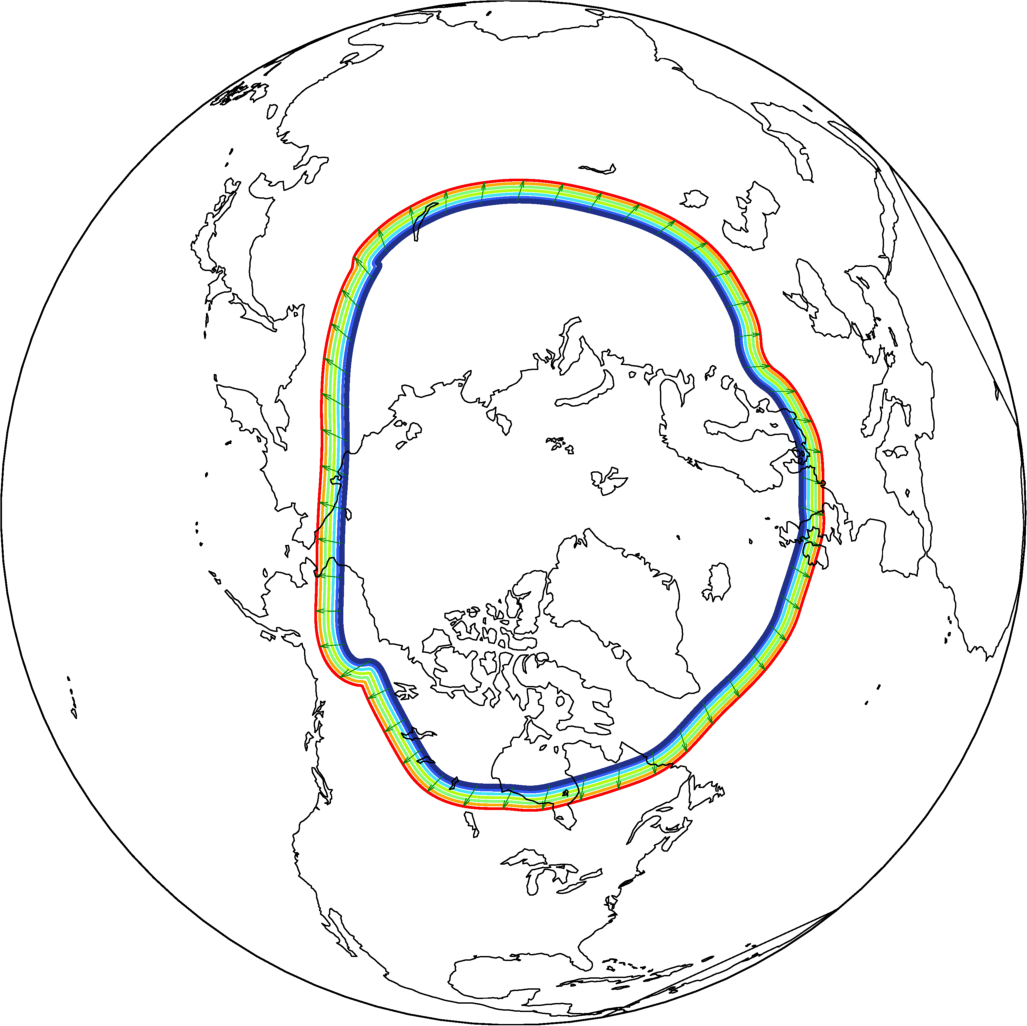}\label{fig:600K_bands}}\ 
	\subfloat[]{\includegraphics[width=0.31\columnwidth]{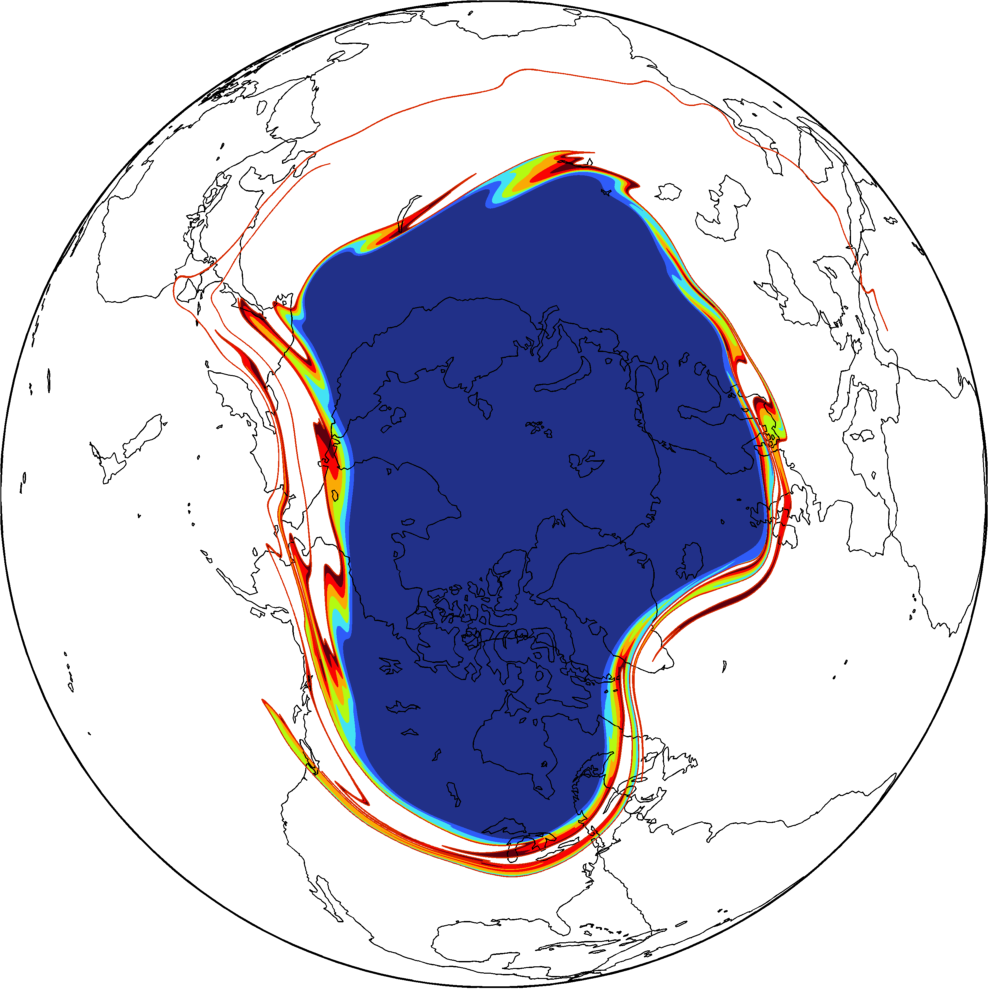}\label{fig:475K_optimal}}\  \subfloat[]{\includegraphics[width=0.31\columnwidth]{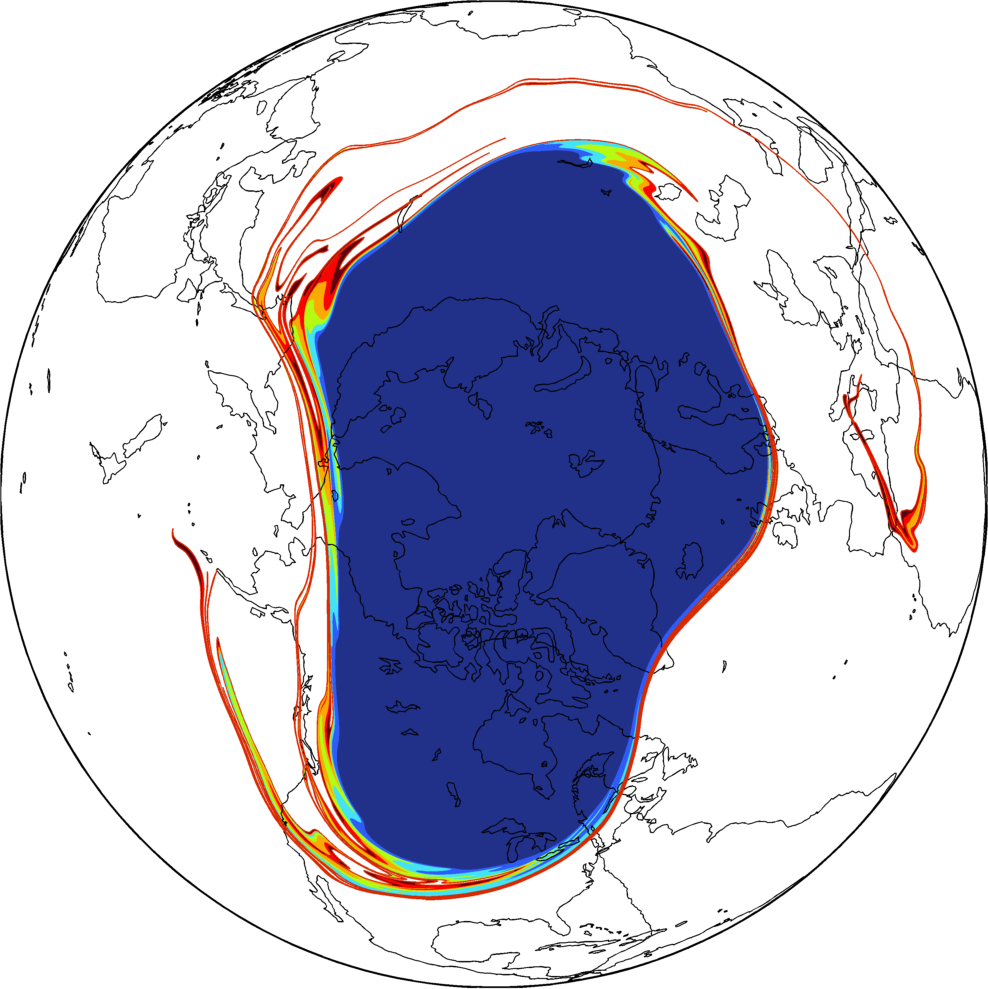}\label{fig:530K_optimal}}\\
	
	\subfloat[]{\includegraphics[width=0.31\columnwidth]{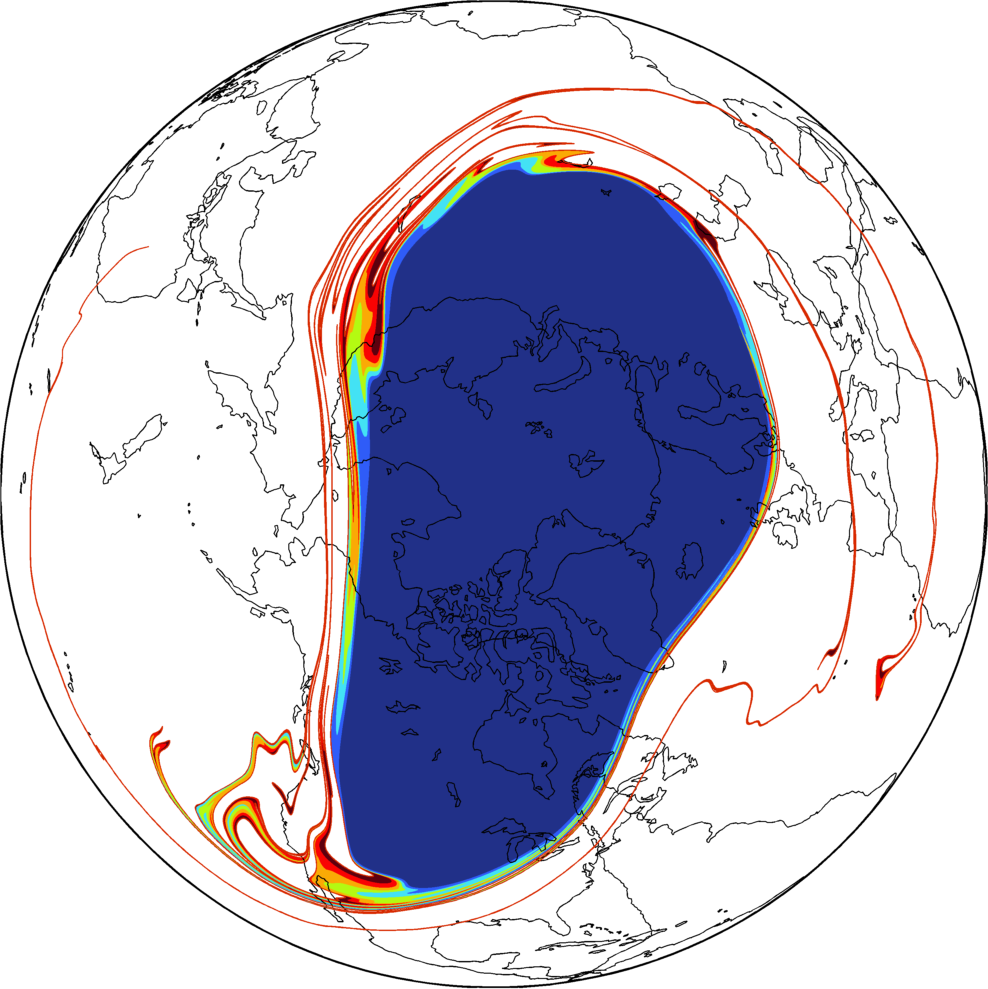}\label{fig:6000K_bands}}\ 
	\subfloat[]{\includegraphics[width=0.31\columnwidth]{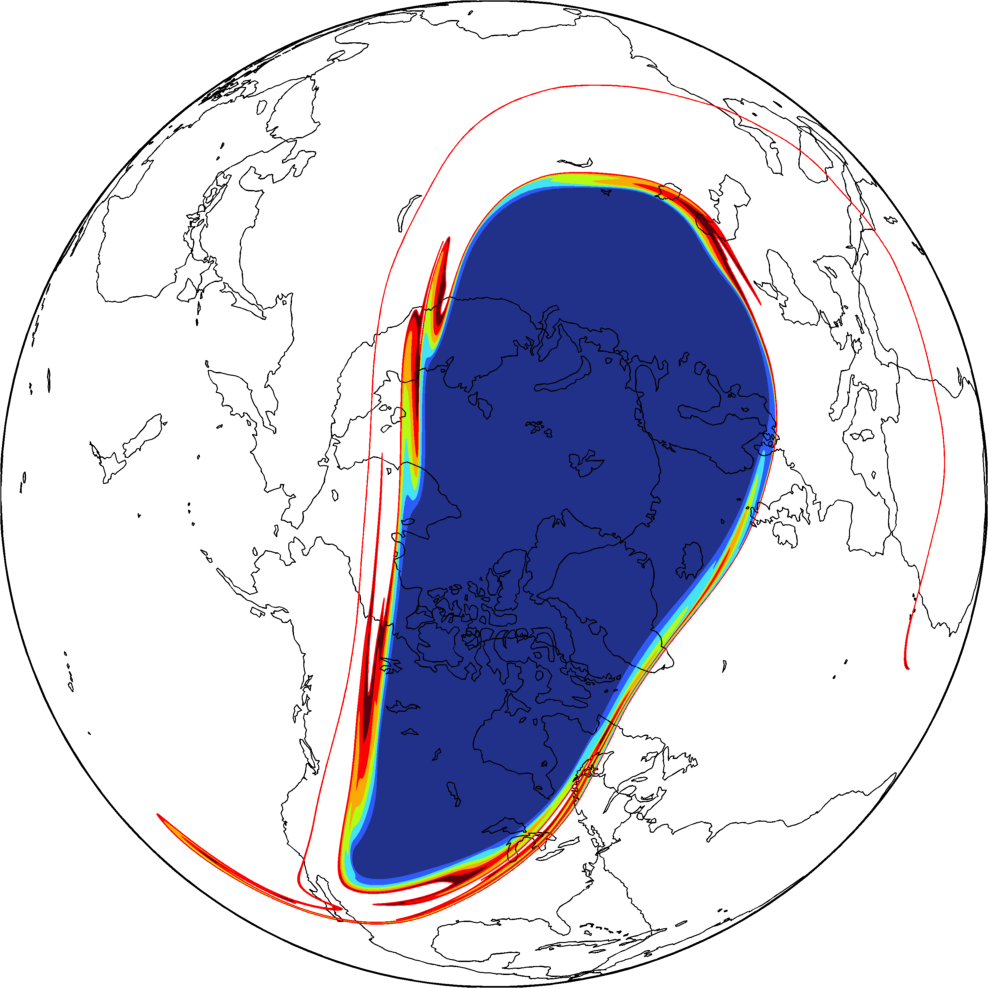}\label{fig:700K_optimal}}\  \subfloat[]{\includegraphics[width=0.31\columnwidth]{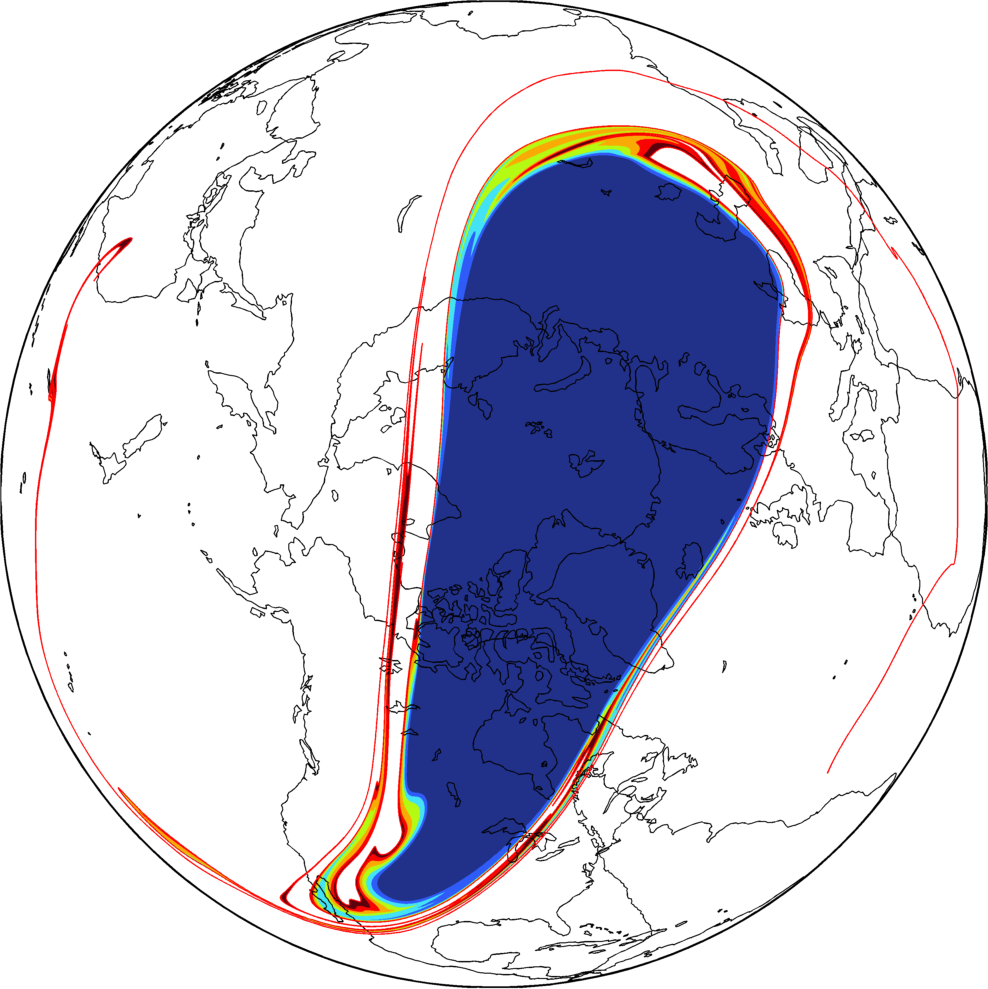}\label{fig:850K_optimal}}\\
	
	\caption{(a) The position of the geodesic vortex edge (blue) on the 600K isentropic surface on 28th Dec. 2013, along with its outward perturbations. Advected images of the geodesic vortex edge and its perturbations on $7^{th}$ Jan. $2014$, for isentropic surfaces (b) $475$K, (c) $530$K, (d) $600$K, (e) $700$K and $850$K.}
	\label{fig:optimality isens}
\end{figure}

 Figure \ref{fig:spherical optimality} shows a combined visualization of the initial positions of the geodesic vortex (blue) together with their perturbations (Fig. \ref{fig:view1}), and their evolved positions (Fig. \ref{fig:view2}) for all isentropic surfaces. A video showing the complete advection sequence over the 10-day time window is available here \href{https://www.dropbox.com/s/4yve97g87vemxol/Figure5MultimediaView.mp4?dl=0}{Video 2}.

\begin{figure}[h]
	\hfill{}\subfloat[]{\includegraphics[width=0.36\columnwidth]{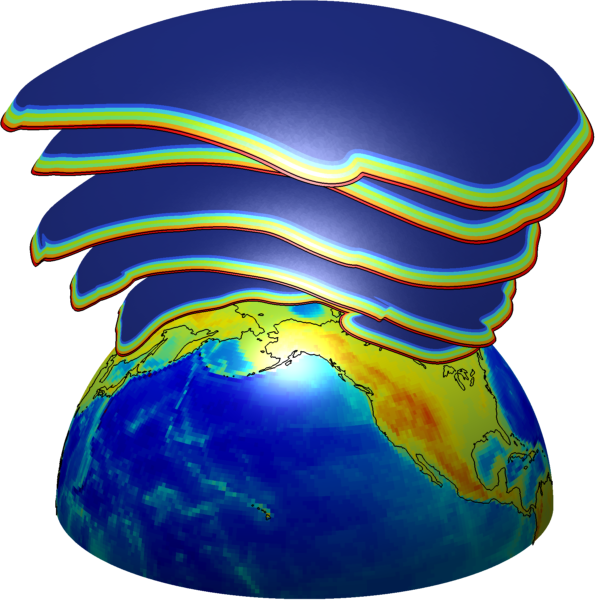}\label{fig:view1}}\hfill{}
	\subfloat[]{\includegraphics[width=0.5\columnwidth]{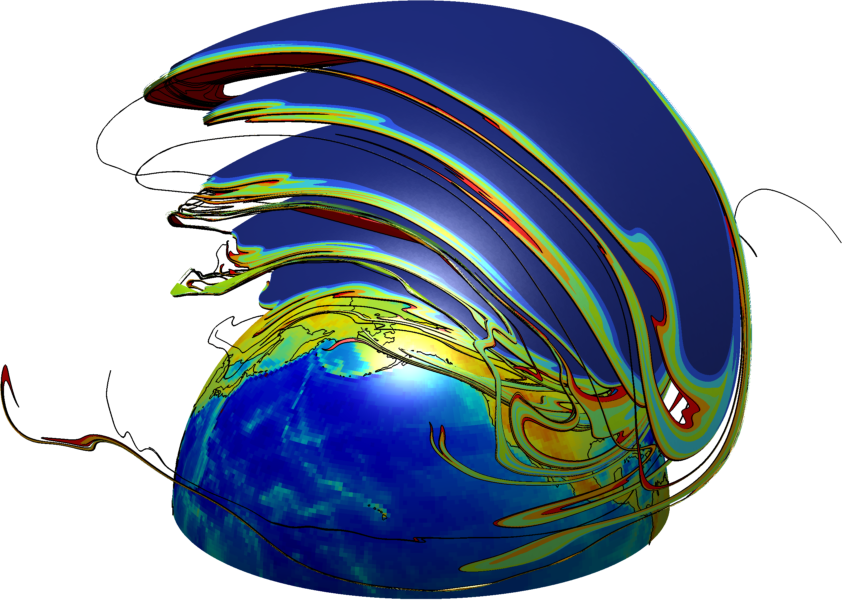}\label{fig:view2}}\hfill{}
	\caption{(a) Geodesic vortex (blue) along with its perturbations on $28^{th}$ Dec. $2013$ on different isentropic surfaces. (b) Advected position of the geodesic vortex and its perturbations on $7{th}$ Jan. $2014$. (Figures not to scale along the radial direction). This 3D visualization shows that the perturbations to the geodesic vortex edge undergo substantial filamentation consistently across all isentropic surfaces. A video showing the complete advection sequence over the 10-days time window is available here \href{https://www.dropbox.com/s/4yve97g87vemxol/Figure5MultimediaView.mp4?dl=0}{Video 2}.}
	\label{fig:spherical optimality}
\end{figure}

\par 
The main vortex-surf zone distinction \cite{mcintrye1983} divides the polar vortex into a coherent vortical air mass, and its weakly coherent surrounding air, which gets eroded by planetary wave-breaking. Thus, ideally, the polar vortex edge should enclose the coherent vortical air mass optimally. Such an optimality necessitates that the immediate exterior of the vortex edge, being a part of the surf zone, exhibit substantial advective mixing with the tropical air. This is because Rossby wave breaking irreversibly erodes the polar vortex, with long filaments of stratospheric air getting pulled off the surf zone, and into the tropical regions \cite{mcintrye1983,mcintyre1984}. In Fig. \ref{fig:spherical optimality}, we observe precisely this behavior in the immediate vicinity of the geodesic vortex edge on each isentropic layer.\par

\subsection{Elliptic LCS and the Nash-method}

 Here we compare the geodesic polar vortex edge with the one obtained using the Nash-method, i.e., the PV isoline possessing the maximum gradient of PV with respect to the equivalent latitude \cite{nash1996}. PV contours are frequently used because in the idealized case of adiabatic and inviscid flows, isentropic PV is conserved, with strong PV gradients generating a restoring force inhibiting meridional transport \cite{vallis}. However, even under these idealistic assumptions, the vortex edge returned by the Nash-method is frame-dependent and non-material, and hence a priori unsuitable for a self-consistent detection of coherent transport barriers.\par

Despite our analysis being purely kinematic, we obtain an overall qualitative agreement between the geodesic vortex boundary and the ones obtained by the Nash-method. However, the Nash-method fails to capture the optimal vortex edge accurately, sometimes underestimating (Fig. \ref{fig:475K_PV_89}) and sometimes overestimating (Fig. \ref{fig:700K_PV_89}) it. \textcolor{black}{The blue areas in Fig. \ref{fig:475K_optimal} and Fig. \ref{fig:700K_optimal} show the advected images of the geodesic vortices in Fig. \ref{fig:475K_PV_89} and Fig. \ref{fig:700K_PV_89}, respectively. Invariably, these areas remain coherent without mixing with warmer air at lower latitudes, as opposed to their slight perturbations. This shows that the outermost materially coherent boundary can either enclose or be enclosed by the one indicated by the Nash method.} Although PV generally decreases with increasing distance from the poles, often small pockets of high PV appear far away from the polar region. Hence, without further filtering, the vortex edge identified by the Nash-method could also include small patches of tropical air (cf. Fig. \ref{fig:475K_PV_89}).

\begin{figure}[h]
	\hfill{}\subfloat[]{\includegraphics[width=0.36\columnwidth]{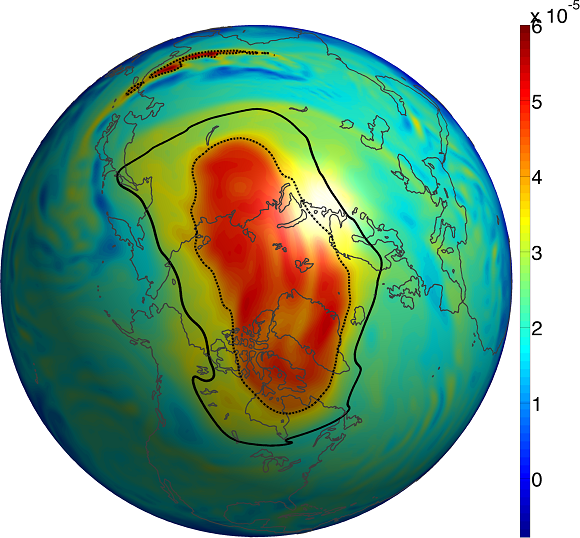}\label{fig:475K_PV_89}}\hfill{}
	\subfloat[]{\includegraphics[width=0.36\columnwidth]{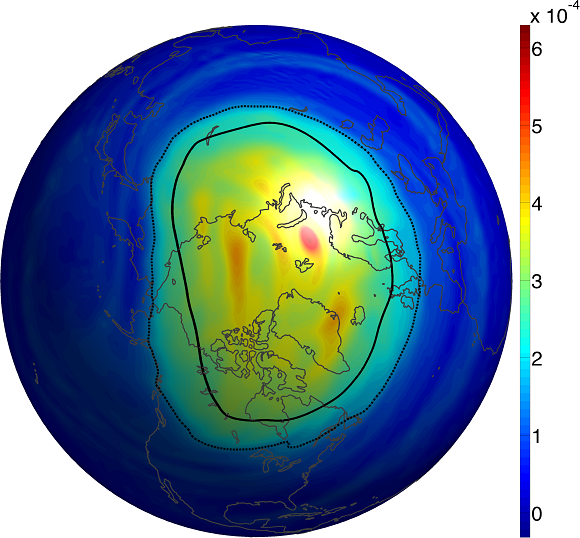}\label{fig:700K_PV_89}}\hfill{}
	\caption{Potential vorticity on (a) the $475$K and (b) the $700$K isentropic surfaces in units of $(K\ m^2)/(kg\ s)$ on $28^{th}$ Dec 2013. The Nash-edge, i.e., the PV contour with the highest gradient of PV with respect to the equivalent latitude, is shown by dashed curves; the geodesic vortex boundary is shown by thick black curves. Small pockets of high PV air are also enclosed by the Nash-edge in Fig. (a). The evolution of the Nash-edge and the geodesic vortex edge during the 10-days time window is available here \href{https://www.dropbox.com/s/7aatri6iq6kh5qw/Figure6MultimediaView.mp4?dl=0}{Video 3}.}
	\label{fig:PV isen}
\end{figure}

 Given its Eulerian (non-material) nature, the Nash-edge evolves discontinuously, with visible jumps in position and shape over time. In contrast, the geodesic vortex edge we extract evolves smoothly because of its material nature. A video comparing the geodesic vortex boundary with the Nash-edge is available here \href{https://www.dropbox.com/s/7aatri6iq6kh5qw/Figure6MultimediaView.mp4?dl=0}{Video 3}. The video shows that sometimes the jumps in the vortex edge identified by the Nash-method are dramatic and include high-value PV arms, which is inconsistent with how the main vortex is envisioned (cf. \textit{Remark 1}).\par
 A video with the three-dimensional evolution of the Nash-edge computed over different isentropic surfaces is available here \href{https://www.dropbox.com/s/h3lzifkmxs50ug3/Figure6MultimediaView_3D.mp4?dl=0}{Video 4} (to be compared with \href{https://www.dropbox.com/s/9d1ykd9egktry0r/Figure3MultimediaView.mp4?dl=0}{Video 1}). Video 4 shows substantial jumps in the bounding surface from one time step to another, due to the non-material nature of the vortex boundaries returned by this method.

\subsection{Elliptic LCSs and the FTLE field}

Backward-time FTLE ridges are popular diagnostics for locating generalized unstable manifolds \cite{shadden2012,yeates2012}. Although the FTLE might incorrectly identify even finite-time generalized stable and unstable manifolds \cite{review,branicki2010,karrasch2015}, it is still used as a diagnostic for locating vortex-type structures, believed to be marked by low FTLE values surrounded by FTLE ridges. In addition to all these caveats, out of the numerous FTLE ridges, there is no clear strategy to filter and extract them as parametrized curves.\par

Figure \ref{fig:FTLE isens} shows the geodesic vortex edge (black) on the $7^{th}$ Jan. 2014 on two different isentropic levels, along with the corresponding backward-time FTLE fields. On each isentropic surface, the FTLE field has a complex filamentary structure, and such complexity increases in surfaces at lower altitudes since the atmospheric motion becomes more chaotic close to the tropopause. Specifically, Fig. \ref{fig:530K_FTLE} corresponds to $475$K, and Fig. \ref{fig:475K_FTLE} to $530$K which is closer to the tropopause, and have numerous strands of high FTLE values emanating towards the equator.\par 

\begin{figure}[h]
	\hfill{}\subfloat[]{\includegraphics[width=0.36\columnwidth]{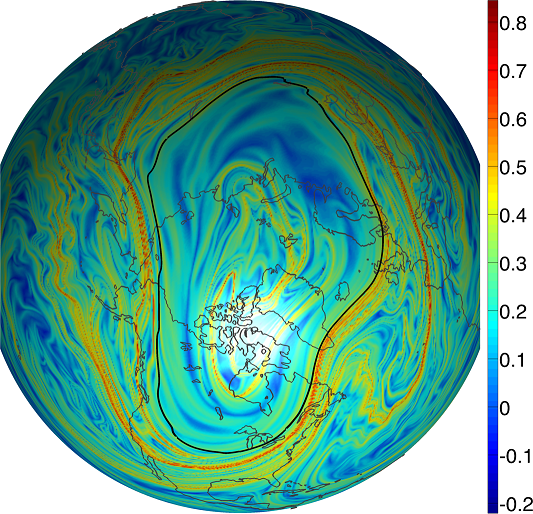}\label{fig:530K_FTLE}}\hfill{}
	\subfloat[]{\includegraphics[width=0.36\columnwidth]{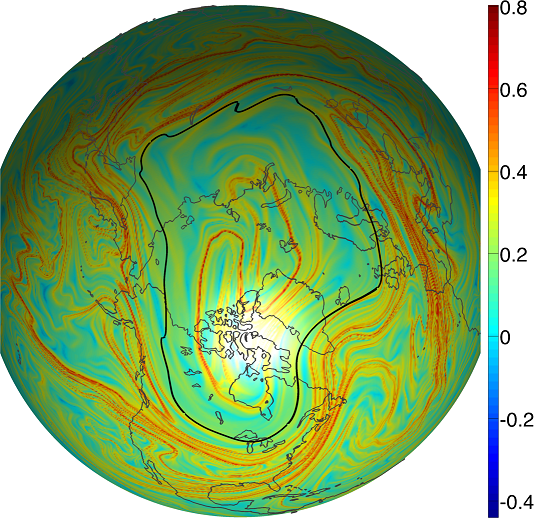}\label{fig:475K_FTLE}}\hfill{}
	\caption{Positions of geodesic vortex boundaries (thick black curves) on $7^{th}$ Jan. 2014, plotted over the backward-time FTLE field on isentropic surfaces (a) $530$K and (b) $475$K.}
	\label{fig:FTLE isens}
\end{figure}

Figure \ref{fig:FTLE isens} shows that parts of some FTLE ridges approximate the optimal location of the polar vortex boundary, but there is no clearly defined, single FTLE ridge marking this boundary. \textcolor{black}{Analogous results would be obtained by computing FTLE over non-Euclidean manifolds with the method developed by Lekien and Ross \cite{lekien2010} (cf. Fig. 9 in \cite{lekien2010})}. Similar conclusions about the forward-time FSLE were been made in \cite{joseph2002}, with the corresponding FSLE ridges penetrating the surf zone and forming a highly mixing `stochastic layer', rather than identifying a coherent transport barrier. In contrast, the geodesic vortex boundary returns the exact location of the coherent vortex edge, as we demonstrated by actual material advection in Section \ref{subsubsection:optimality}.

\subsection{Elliptic LCS and the ozone concentration field}

The location and shape of the geodesic vortex edge is consistent with the ozone concentration (cf. Fig. \ref{fig:ozone}), in agreement with the chemical composition expected within the main vortex. Ozone concentration behaves approximately like a passive tracer in the stratosphere over several weeks \cite{middle_atmosphere}, and thus evolves with the vortex. Ozone hole formation is anti-correlated to the planetary wave activity \cite{bodekar1995,shindell1997}, and thus depends on the strength of the meridional transport barrier enclosing the polar vortex. For example, it has been shown that the polar night jet, a barrier to stratospheric meridional transport of passive tracers such as ozone, accounts for the sharp boundary of the Antarctic ozone hole \cite{rypina2006}. Despite the lower subsistence of the Arctic vortex compared to the Antractic vortex \cite{waugh1999}, the geodesic vortex edge shadows closely the boundary separating regions of contrasting ozone concentrations over the time window we analyze at every isentropic layer. 

\begin{figure}[h]
	\hfill{}\subfloat[]{\includegraphics[width=0.36\columnwidth]{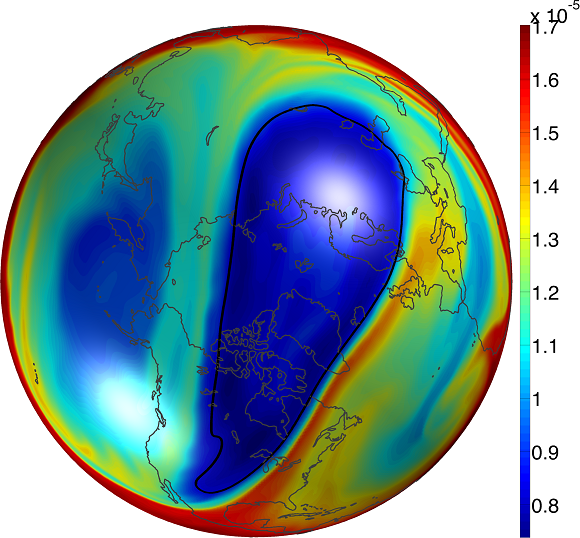}\label{fig:ozone_850K_99}}\hfill{}
	\subfloat[]{\includegraphics[width=0.36\columnwidth]{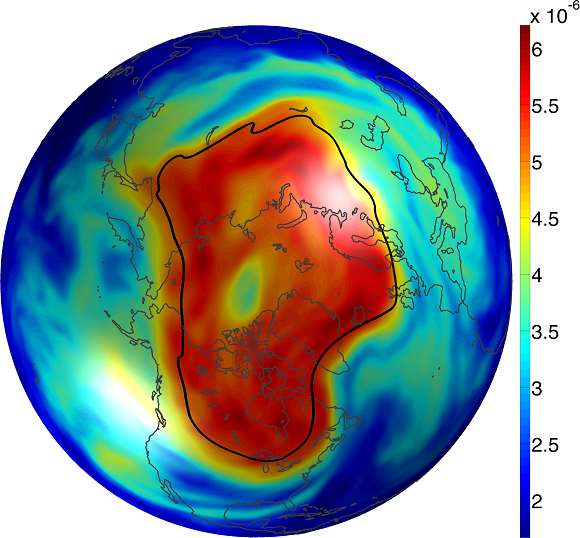}\label{fig:ozone_445K_99}}\hfill{}
	\caption{Nearby sharp contrast in the ozone concentration is observed along the geodesic vortex boundary of each isentropic surface. Shown here are the ozone mass mixing ratios in units of $(kg)/(kg)$ on (a) the $850$K and (b) the $475$K isentropic surfaces on $7^{th}$ Jan. $2014$.}
	\label{fig:ozone}
\end{figure}

As an illustrative example, Figs. \ref{fig:ozone_850K_99}-\ref{fig:ozone_445K_99} show the vortex edge (black) on $7^{th}$ Jan $2014$ on $850$K and $475$K, along with their corresponding ozone concentration. Note the relatively lower and higher ozone concentrations in the polar vortex  compared with the surrounding air, are a result of the Brewer-Dobson circulation \cite{mohanakumar2008}. In this mechanism, ozone created in the tropical stratosphere is transported polewards in the lower stratosphere. 

\subsection{Elliptic LCS and the temperature field}
Temperature does not behave like a passive tracer on isentropic surfaces, hence it is not expected to advect as air particles. However, curiously, we generally find that geodesic vortex boundaries on isentropic surfaces turn out to match remarkably well with the temperature fields available on isobaric surfaces roughly 10 km lower in altitude. As an illustration, we show that the geodesic boundaries of the $850$K and $700$K isentropic surfaces match well with the temperature fields of the $30$hPa and $50$hPa isobaric surfaces (cf. Fig. \ref{fig:iso_isen_temps}). Note that we use temperature fields on isobaric surfaces available at \cite{japmet} because they are not directly available on isentropic surfaces in the ECMWF database. 

\begin{figure}[h]
	\hfill{}\subfloat[]{\includegraphics[width=0.36\columnwidth]{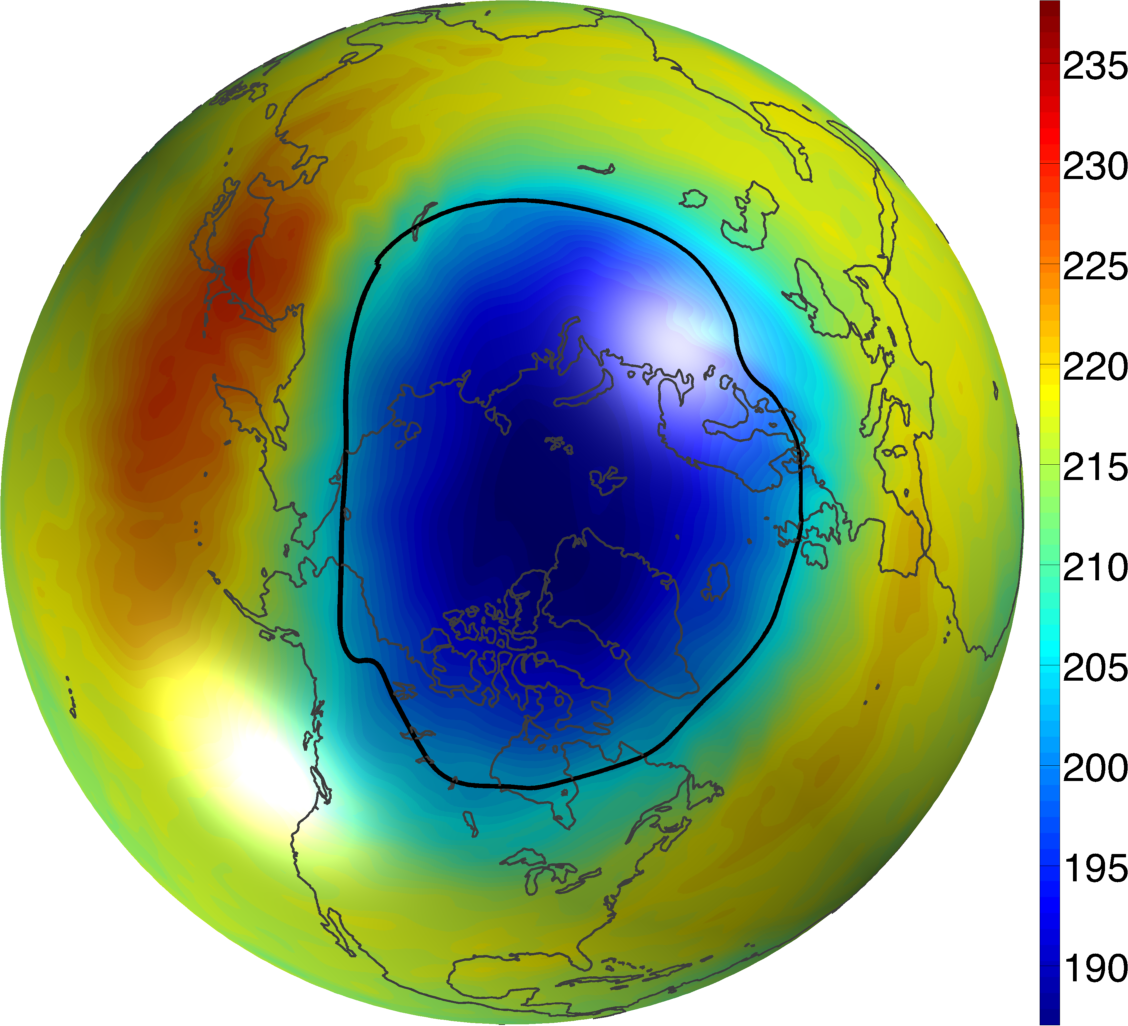}\label{fig:mix_temps_850K_89}}\hfill{}
	\subfloat[]{\includegraphics[width=0.36\columnwidth]{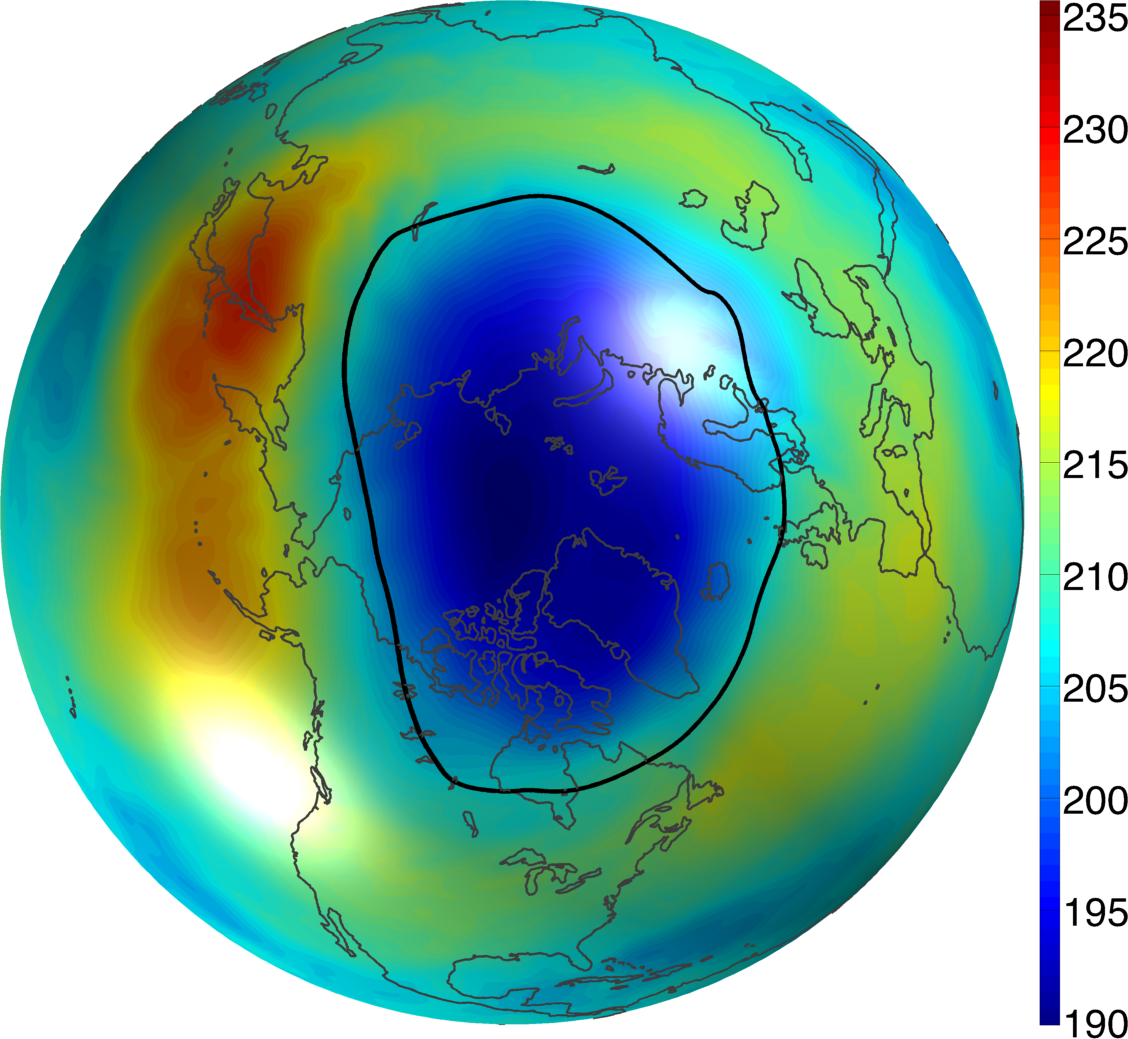}\label{fig:mix_temps_7000K_89}}\hfill{}
	\caption{Temperature fields (in Kelvin) of (a) $30$hPa and (b) $50$hPa isobaric surfaces and geodesic vortex boundaries on (a) $850$K and (b) $700$K isentropic surfaces on $28^{th}$ Dec. $2013$.}
	\label{fig:iso_isen_temps}
\end{figure}

Despite their purely kinematic nature, LCSs show a close match with instantaneous temperature and ozone concentration fields, bringing further evidence that the geodesic vortex boundary correctly identifies the physically observable polar vortex edge.

\subsection{Vortex shrinkage and vertical motion inside the vortex}
The assumption of adiabatic and \textcolor{black}{frictionless} motion implies the restriction of atmospheric dynamics to quasi-horizontal isentropic surfaces. In reality, however, radiative heating and cooling effects invariably arise, resulting in cross-isentropic vertical motion inside the vortex. The well-known Brewer-Dobson circulation  \textcolor{black}{\cite{vallis}} posits a vertically downward and horizontally poleward movement of air in the winter stratosphere. Cross-isentropic vertical motion, in the form of diabatic descent, has been calculated in various studies. Based on diabatic heating computations using a radiation model, Rosenfield et al.
 \cite{rosenfield1994} conclude that the amount of descent in the middle and upper stratosphere is substantially larger than in the lower stratosphere. Mankin et al. \cite{mankin1990} and  Shroeberl et al. \cite{schoeberl1992} also conclude that the magnitude of descent increases with altitude.\par
The shrinkage of the vortex area on a cross-sectional surface is an indicator of the magnitude of the vertical motion inside the full three-dimensional vortex. Over the $10$-day time window, the area enclosed by the geodesic vortex edge decreased on each of the five isentropic surfaces (Fig. \ref{fig:areas}). Moreover, the magnitude of this shrinkage is consistently higher for surfaces with higher potential temperatures (which are at higher altitudes; cf. Fig. \ref{fig:all shrinkage}). The shrinkage of the vortex on the $850$ K isentropic surface is substantially higher than the rest, in agreement with the diabatic descent trend observed in \cite{rosenfield1994}.

\begin{figure}[H]
	\subfloat[]{\includegraphics[width=0.4\columnwidth]{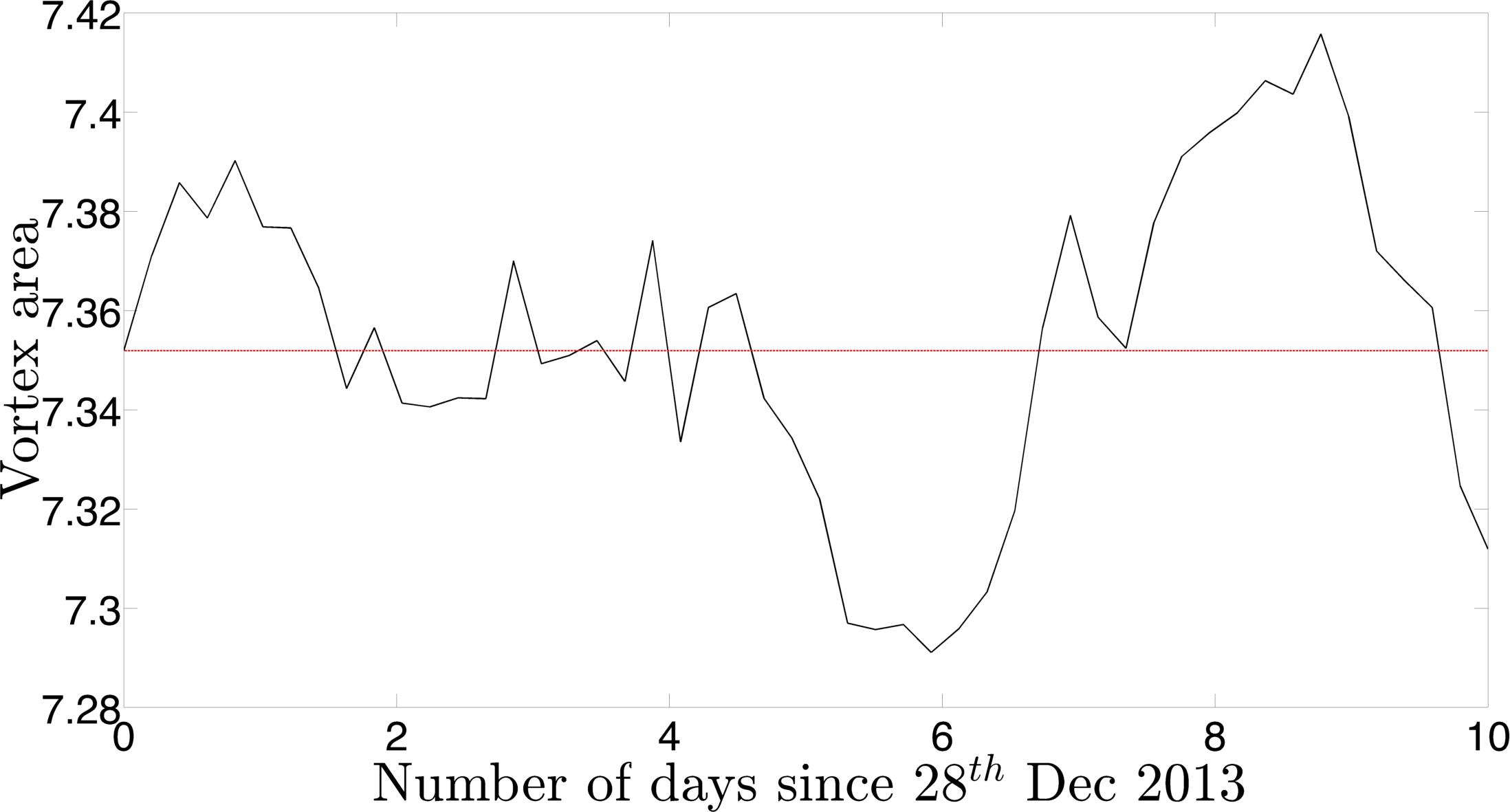}\label{fig:475_vortex_shrinkage}}\hfill{}
	\subfloat[]{\includegraphics[width=0.4\columnwidth]{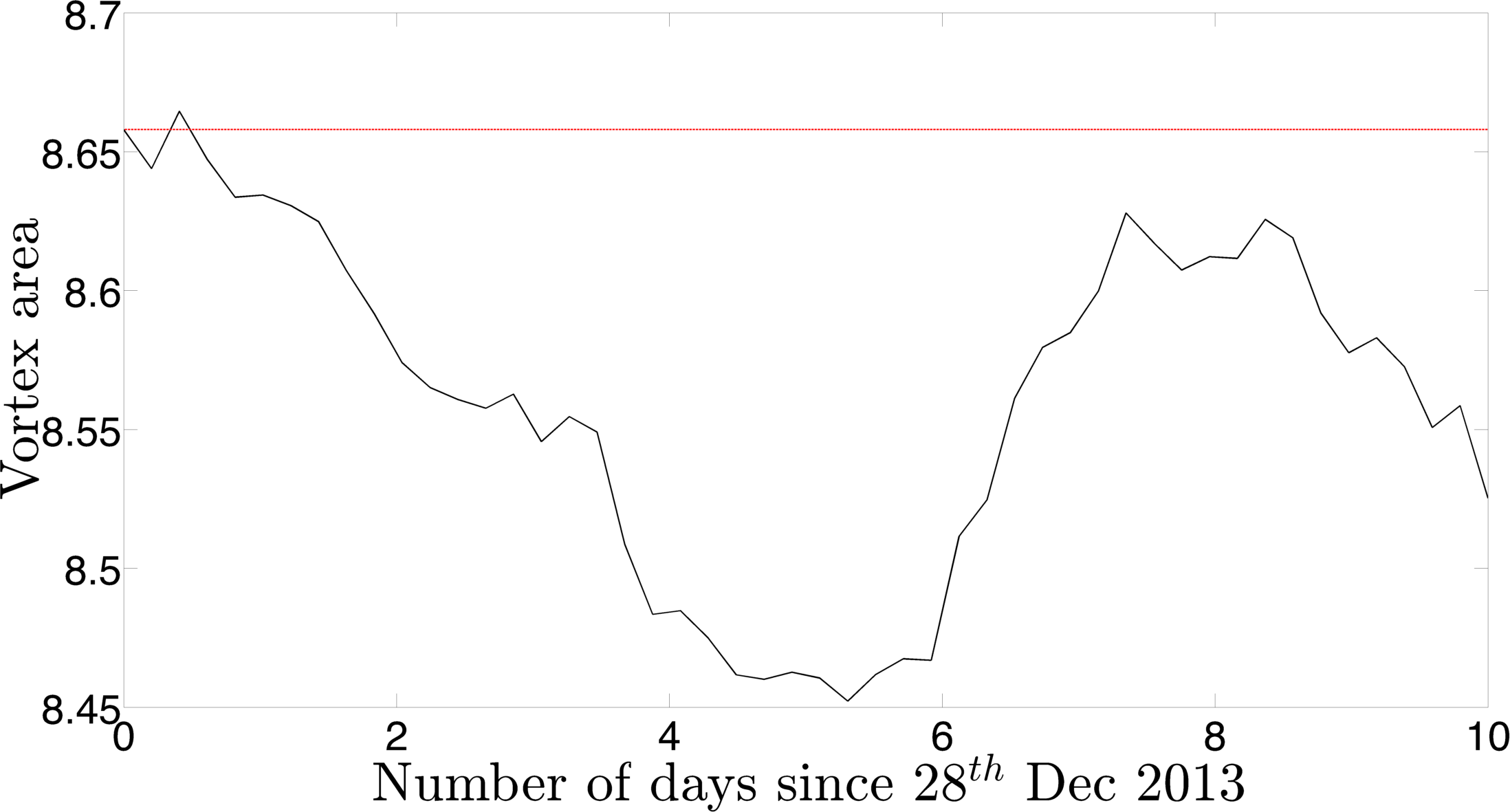}\label{fig:530_vortex_shrinkage}}\\
	\subfloat[]{\includegraphics[width=0.4\columnwidth]{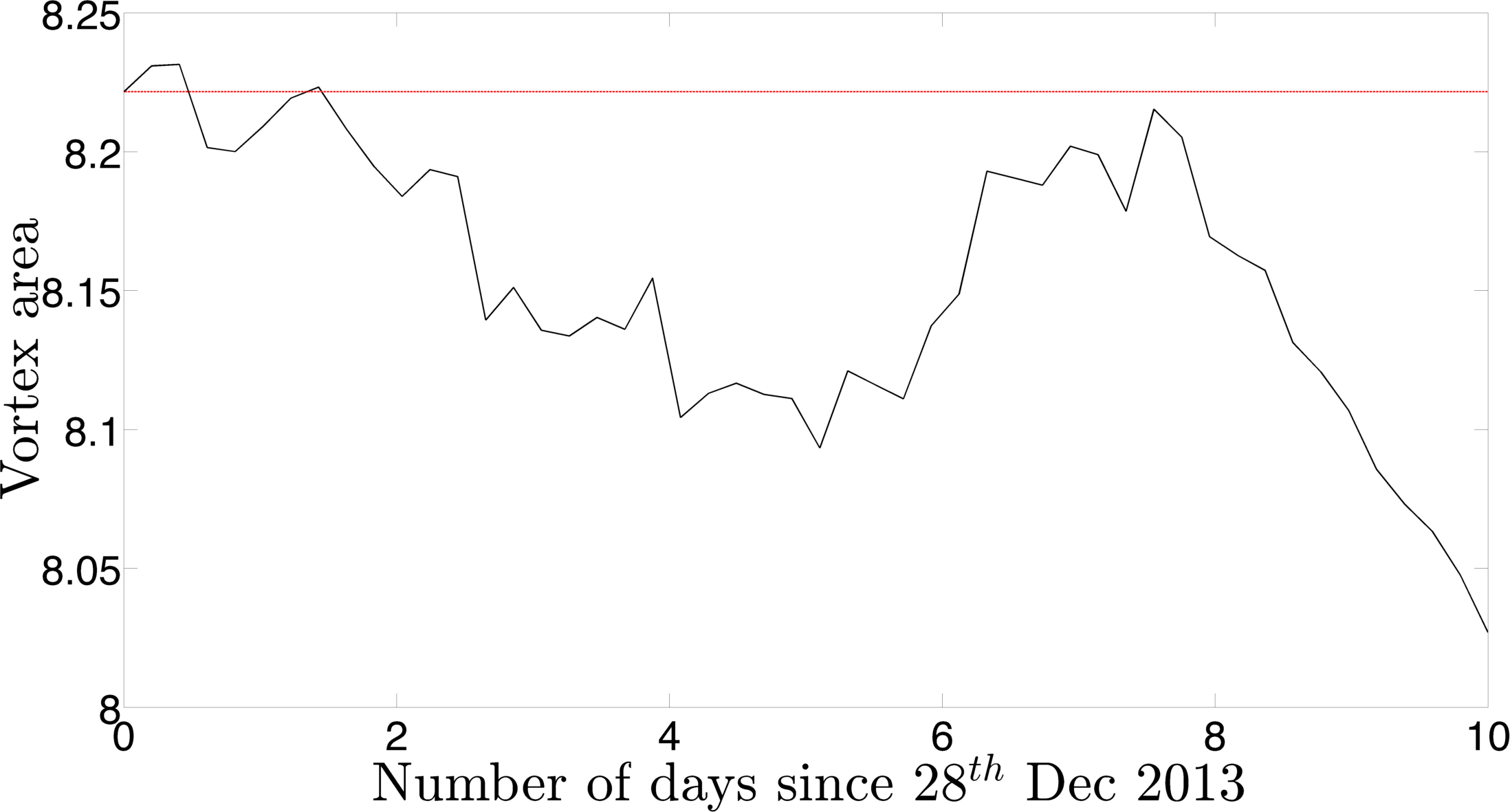}\label{fig:600_vortex_shrinkage}}\hfill{}
	\subfloat[]{\includegraphics[width=0.4\columnwidth]{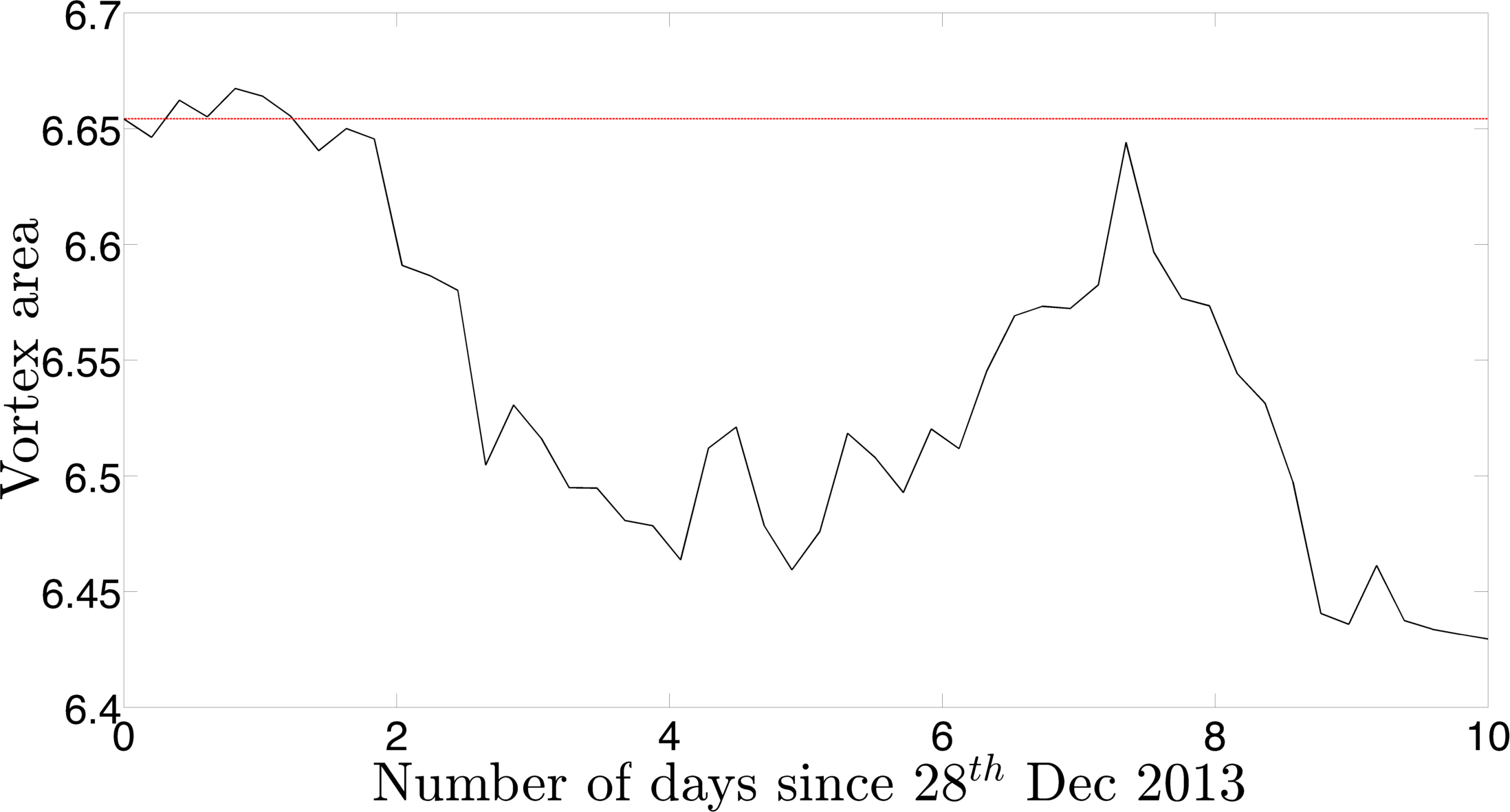}\label{fig:700_vortex_shrinkage}}\\
	\subfloat[]{\includegraphics[width=0.4\columnwidth]{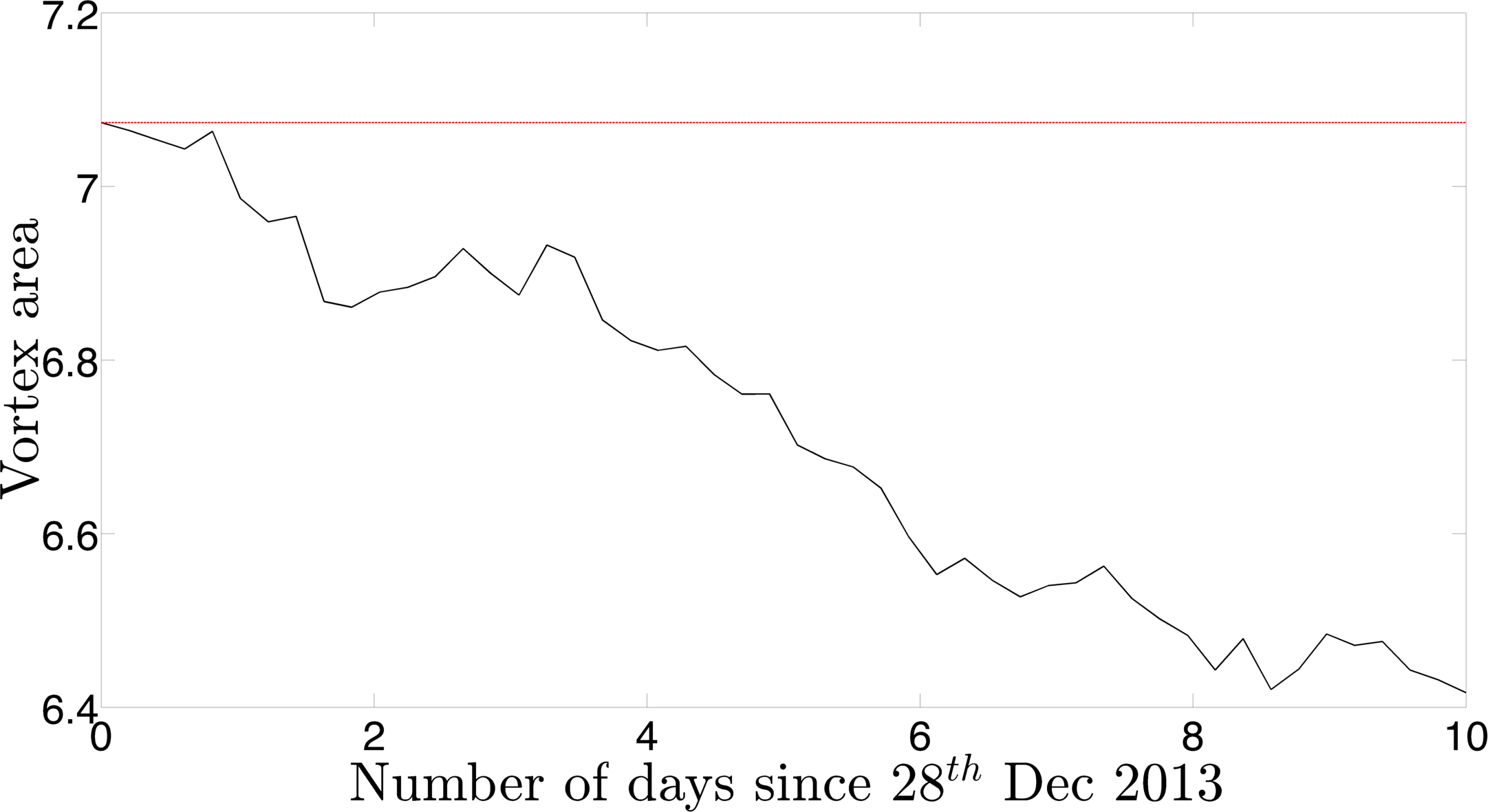}\label{fig:850_vortex_shrinkage}}\hfill{}
	\subfloat[]{\includegraphics[width=0.4\columnwidth]{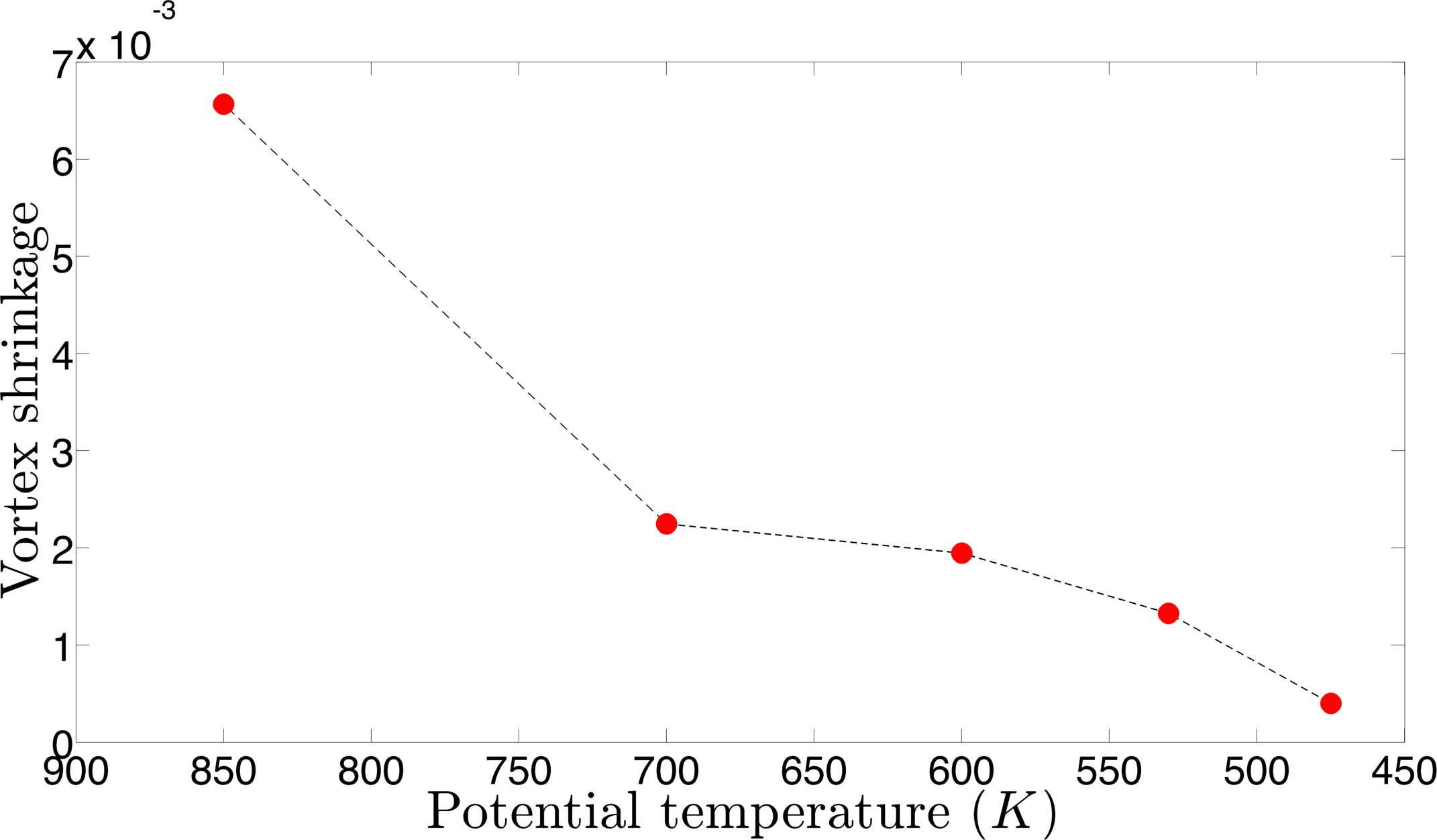}\label{fig:all shrinkage}}\\
	\caption{(a-e) The geodesic vortex area evolution with time, on (a) $475$K, (b) $530$K, (c) $600$K, (d) $700$K and (e) $850$K isentropic surfaces in the units of percentage of Earth's surface area. (f) 10-days vortex area shrinkage in units of percentage of Earth's surface area as a function of potential temperature.}
	\label{fig:areas}
\end{figure}

\section{Conclusions}
The two polar vortices are dominant dynamical features of stratospheric circulation. During fall and winter, they exemplify coherent vortical motion, with a delineating transport barrier commonly referred to as the `vortex edge', outside which lies the highly mixing `surf zone'. Recent interest in the polar vortices has been motivated by the central role they play in the ozone hole formation \cite{WMO1999,McIntyre1995}, as well as by their influence over tropospheric weather \cite{thompson1998,thompson2000}. \par
Using the recently developed geodesic theory of Lagrangian coherent structures \cite{review,serra2016efficient}, we have computed geodesic vortex boundaries on various isentropic surfaces during the late December 2013 and early January 2014, when an exceptional cold wave was recorded in the Northeastern US. Geodesic LCS theory enables us to identify the polar vortex edge as a smooth parametrized material surface. In comparison to other diagnostics used for the vortex edge identification, the geodesic LCS method uniquely identifies a materially optimal (i.e., maximal and non-filamenting) vortex boundary, dividing the main vortex from the surf zone. We verify this optimality by actual material advection of the geodesic vortex edge and its normal perturbations. Remarkably, we find that even slightly normally perturbed surfaces exhibit substantial advective mixing with the tropical air, while the geodesic edge remains perfectly coherent, conforming to the original idea for a vortex edge. We find that the polar vortex is initially roughly symmetrically placed in late December 2013, while it deforms towards the Northeast coast of the US in early January 2014, consistent with the severe cold registered over this period.\par 

Isentropic PV was instrumental in realizing the `main vortex-surf zone' \cite{mcintyre1984,juckes1987,mcintrye1983} distinction in the stratospheric polar vortex. A popular PV-based method posits the vortex edge as the PV contour with the highest PV gradient with respect to equivalent latitude \cite{nash1996}. We have observed that despite its simplicity, this method sometimes underestimates while at other times overestimates the extent of the coherent vortex core. Furthermore, due to their Eulerian nature, PV-based methods are inherently sub-optimal for material assessments. The geodesic LCS theory used here is purely kinematic, and hence model-independent. Not only it is immune to model-dependent fallibility of diagnostics, such as PV, but it is also objective (i.e. frame-independent). Despite its kinematic nature, the method renders a vortex edge that closely match sharp gradients in the temperature field and ozone concentration.\par
Using the shrinkage of the coherent vortical area on different isentropic surfaces, we obtain a good agreement with the trend of increasing diabatic descent with increasing altitude, observed also in several other studies \cite{rosenfield1994,mankin1990,schoeberl1992}.
\section*{Acknowledgments}
We acknowledge helpful discussions with Mohammad Farazmand and Maria Josefina Olascoaga, and Pak Wai Chan for pointing out the source of isobaric surface data.

\bibliographystyle{plain}
\bibliography{PVDatabase_v2}

\end{document}